\documentclass[preprint2]{aastex61}
\usepackage{amsmath,empheq}
\usepackage{natbib}
\usepackage{hyperref}
\usepackage{color}

\newcommand\tq[3]{#1\substack{+#2 \\ -#3}}

\newcommand\hiresult{$\tq{1.24}{0.23}{0.20}\ $}
\newcommand\htd{$\tq{0.94}{0.20}{0.18}$}
\newcommand\htf{ $\tq{0.29}{0.14}{0.15}$}

\newcommand\gresult{$\tq{1.50}{0.19}{0.18}\ $}
\newcommand\gtd{$\tq{0.69}{0.13}{0.13}$}
\newcommand\gtf{$\tq{0.80}{0.15}{0.18}$}

\newcommand\simhiresult{\sim1.2}
\newcommand\simgresult{\sim1.5}

\newcommand{\rpm}{\raisebox{.2ex}{$\scriptstyle\pm$}}

\mathchardef\mhyphen="2D
\shorttitle{The Evolution of Environmental Quenching Timescales to $z\sim1.6$}
\shortauthors{Foltz et al.}

\begin{document}

\title{The Evolution of Environmental Quenching Timescales to $z\sim1.6$: Evidence for Dynamically-Driven Quenching of the Cluster Galaxy Population}

\author{R. Foltz}
\affil{Department of Physics \& Astronomy, University of California Riverside, 900 University Avenue, Riverside, CA 92521}
\email{scrudgey@gmail.com}

\author{G. Wilson}
\affil{Department of Physics \& Astronomy, University of California Riverside, 900 University Avenue, Riverside, CA 92521}
\email{gillian.wilson@ucr.edu}

\author{A. Muzzin}
\affil{Department of Physics and Astronomy, York University, 4700 Keele St., Toronto, ON MJ3 1P3, Canada}

\author{M. C. Cooper}
\affil{Department of Physics \& Astronomy, 4129 Reines Hall, University of California, Irvine, CA 92697}

\author{J. Nantais}
\affil{Departamento de Ciencias Físicas, Universidad Andrés Bello, Santiago, Chile}

\author{R.F.J. van der Burg}
\affil{European Southern Observatory, Karl-Schwarzschild-Str. 2, 85748, Garching, Germany}

\author{P. Cerulo}
\affil{Department of Astronomy, Universidad de Concepci\'on, Casilla 160-C, Concepci\'on, Chile}


\author{J. Chan}
\affil{Department of Physics \& Astronomy, University of California Riverside, 900 University Avenue, Riverside, CA 92521}

\author{S. P. Fillingham}
\affil{Department of Physics \& Astronomy, 4129 Reines Hall, University of California, Irvine, CA 92697}

\author{J. Surace}
\affil{Infrared Processing and Analysis Center, California Institute of Technology, Pasadena, CA, 91125}

\author{T. Webb}
\affil{Department of Physics, 3600 rue University, Montreal QC, H3A 2T8, Canada}

\author{A. Noble}
\affil{Department of Astronomy \& Astrophysics, University of Toronto, Ontario M5S 3H4, Canada}

\author{M. Lacy}
\affil{National Radio Astronomy Observatory, Charlottesville, VA 22903}

\author{M. McDonald}
\affil{Massachusetts Institute of Technology, 77 Massachusetts Avenue, Cambridge, MA 02139}

\author{G. Rudnick}
\affil{Department of Physics \& Astronomy, University of Kansas, 1251 Wescoe Hall Drive, Lawrence, KS, 66045}

\author{C. Lidman}
\affil{Australian Astronomical Observatory, PO Box 2915, North Ryde NSW 1670, Australia}

\author{R. Demarco}
\affil{Departamento de Astronom\'ia, Universidad de Concepci\'on, Casilla 160-C, Concepci\'on, Regi\'on del Biob\'io, Chile}

\author{J. Hlavacek-Larrondo}
\affil{Département de Physique, Université de Montréal, Montréal, QC H3C 3J7, Canada}

\author{H.K.C. Yee}
\affil{Department of Astronomy \& Astrophysics, University of Toronto, Ontario M5S 3H4, Canada}

\author{S. Perlmutter}
\affil{Lawrence Berkeley National Laboratory and Center for Particle Astrophysics, U.C. Berkeley}

\author{B. Hayden}
\affil{Lawrence Berkeley National Laboratory and Center for Particle Astrophysics, U.C. Berkeley}

\begin{abstract}
Using a sample of 4 galaxy clusters at $1.35 < z < 1.65$ and 10 galaxy clusters at $0.85 < z < 1.35$, we measure the environmental quenching timescale, $t_Q$, corresponding to the time required after a galaxy is accreted by a cluster for it to fully cease star formation.
Cluster members are selected by a photometric-redshift criterion, and categorized as star-forming, quiescent, or intermediate according to their dust-corrected rest-frame colors and magnitudes.
We employ a ``delayed-then-rapid" quenching model that relates a simulated cluster mass accretion rate to the observed numbers of each type of galaxy in the cluster to constrain $t_Q$.
For galaxies of mass $M_* \gtrsim 10^{10.5}~ \mathrm{M}_\odot$, we find a quenching timescale of $t_Q=$ \hiresult Gyr in the $z\sim1.5$ cluster sample, and $t_Q=$ \gresult Gyr at $z\sim1$.
Using values drawn from the literature, we compare the redshift evolution of $t_Q$ to timescales predicted for different physical quenching mechanisms.
We find $t_Q$ to depend on host halo mass such that quenching occurs over faster timescales in clusters relative to groups, suggesting that properties of the host halo are responsible for quenching high-mass galaxies.
Between $z=0$ and $z=1.5$, we find that $t_Q$ evolves faster than the molecular gas depletion timescale and slower than an SFR-outflow timescale, but is consistent with the evolution of the dynamical time.
This suggests that environmental quenching in these galaxies is driven by the motion of satellites relative to the cluster environment, although due to uncertainties in the atomic gas budget at high redshift, we cannot rule out quenching due to simple gas depletion.
\end{abstract}


\section{Introduction}

Galaxies form a bimodal distribution in rest-frame color at $z < 2$ \citep{Strateva:2001aa,Baldry:2004aa,Bell:2004aa,Williams:2009tt}, meaning galaxies can be broadly categorized as either actively star-forming spirals (the ``blue cloud"), or quiescent ellipticals and lenticulars (the ``red-sequence").
Although these populations are roughly equivalent in total stellar mass at $z\sim1$, the quiescent galaxy population has nearly doubled in stellar mass, stellar mass density, and number density over the past $\sim7$ Gyr \citep{Bell:2004aa,Borch:2006aa,Bundy:2006aa,Arnouts:2007aa,Brown:2007aa,Faber:2007aa}.

Meanwhile, a variety of studies at intermediate redshift show that galaxy properties correlate with local environment \citep{Cooper:2006aa,Cooper:2007aa,Quadri:2007aa,Patel:2009aa}, such that groups and clusters contain more quiescent than active galaxies \citep{George:2011aa,Muzzin:2012dw,Presotto:2012aa,Tanaka:2012aa,Nantais:2017aa}.
Moreover, with increasing cluster-centric radius (decreasing time since infall), observations find a relative reduction in the number of quiescent systems \cite[e.g.][]{Presotto:2012aa}.
Together, these results suggest that dense environments shut off (or ``quench") star formation in galaxies —-- a process typically termed ``environmental quenching" \citep{Peng:2010aa}.
Environment has been studied extensively as a driver of galaxy evolution \citep[for a review see][]{Blanton:2009aa}, but the physical mechanism or mechanisms responsible for quenching have yet to be identified, although several candidates have been proposed.

Whatever the underlying cause of quenching, it must disrupt the process by which a galaxy converts cold gas into stars.
As a galaxy forms stars, its cold gas reservoir is replenished as its surrounding hot gas halo cools \citep{Bauermeister:2010aa}.
One possibility is that this gas is directly removed from a galaxy by ram-pressure stripping as it falls at high speed into the hot intra-cluster medium (ICM) of a cluster environment \citep{Gunn:1972aa}.
If the gas is not stripped, then the role of the environment may be simply to prevent the accretion of fresh gas onto the galaxy, causing the galaxy to quench as star formation exhausts the remaining gas reservoir over a gas depletion time.

It is also possible that feedback and outflows play a key role in removing the gas from galaxy halos \citep{McGee:2014aa,Balogh:2016aa}.
In this ``overconsumption" scenario, the depletion of gas is augmented by outflows produced by star formation, either directly through radiation pressure or from subsequent supernovae \citep{McGee:2014aa}.
Quenching then proceeds over an accelerated gas depletion timescale that is inversely proportional to the star formation rate (SFR).

These processes act to quench galaxies over different timescales, and differences between the predictions become more apparent with increasing redshift.
Measuring the evolution in the quenching timescale, $t_Q$, over as large a redshift baseline as possible is therefore a powerful approach to identifying which of the proposed mechanisms discussed above may be primarily responsible for causing the quenching.

In this work, we will measure the quenching timescale in a sample of four galaxy clusters at $z \sim 1.6$, a higher redshift than has been studied previously.
We will use our measurements, together with other measurements at lower redshift drawn from the literature, to investigate the redshift evolution of $t_Q$ compared to model predictions for different quenching mechanisms.

The quenching timescale analysis presented here complements previously published environmental quenching studies carried out by our own group and others.
At $z\sim1.6$, \citet{Noble:2017aa} find that cluster galaxies lie systematically at higher gas fractions and longer depletion timescales than the field scaling relations.
Between $z=1.6$ and $z=0.9$, \citet{Nantais:2016aa,Nantais:2017aa} find a strong evolution in environmental quenching efficiency while, over a similar redshift range, \citet{Cerulo:2016aa,Cerulo:2017aa} report an accelerated build-up of the red-sequence in clusters.
At $z\sim1$, \citet{Muzzin:2012dw} measured a quenching timescale of $\sim1$ Gyr based on an analysis of poststarburst galaxies.
Between $z\sim1$ and $z=0$, \citet{Balogh:2016aa} finds evidence for a change in the dominant environmental quenching mechanism.



The structure of this paper is as follows:
Our data set is described in Section \ref{sec-data}.
In Section \ref{sec-analysis}, we summarize our toy model of environmental quenching, which is described in detail in Appendix \ref{sec-math}.
In Section \ref{sec-results} we report the results of our technique, which we discuss in Section \ref{sec-discussion}.
In Section \ref{sec-conclusion} we summarize our conclusions.

In this work we will assume a standard $\Lambda$CDM cosmology with $H_0 = 70 \mathrm{\ km \cdot s^{-1} \cdot Mpc^{-1}}, \Omega_M = 0.3, \mathrm{\ and\ } \Omega_\Lambda = 0.7$, and a Chabrier IMF \citep{chabrierimf} throughout.
Our magnitudes are reported in the AB system \citep{Oke:1983aa}.

\section{Data}\label{sec-data}

\begin{deluxetable*}{lcccccccc}
\tabletypesize{\scriptsize}
\tablecaption{Description of the $z\sim1.6$ SpARCS cluster sample\label{tbl-clusters}}
\tablewidth{0pt}
\tablehead{
\colhead{Cluster}
& \colhead{R.A.}
& \colhead{Decl.}
& \colhead{z}
& \colhead{Spectroscopy}
& \colhead{Photometry}
& \colhead{Spectra$^\mathrm{a}$}
& \colhead{N$_\mathrm{spec}^\mathrm{b}$}
}
\startdata
SpARCS-J0224 & 02:24:26.33 & -03:23:30.8 & 1.633 & MOSFIRE, FORS2, OzDES & \textit{ugrizYJKs} [3.6] [4.5] [5.8] [8.0] & 187 & 52 \\
SpARCS-J0330 & 03:30:55.87 & -28:42:59.5 & 1.626 & MOSFIRE, FORS2, OzDES & \textit{ugrizYJKs} [3.6] [4.5] [5.8] [8.0] & 535 & 40 \\
SpARCS-J0225 & 02:25:45.55 & -03:55:17.1 & 1.598 & MOSFIRE, FORS2, OzDES & \textit{ugrizYKs} [3.6] [4.5] [5.8] [8.0] & 126 & 22 \\
SpARCS-J0335 & 03:35:03.58 & -29:28:55.6 & 1.369 & FORS2, OzDES & \textit{grizYKs} [3.6] [4.5] [5.8] [8.0] & 81 & 22 \\
\enddata
\tablenotetext{a}{Number of spectra.}
\tablenotetext{b}{Number of spectroscopically-confirmed cluster members.}
\end{deluxetable*}

The galaxy clusters studied in this work were identified using the Stellar Bump Sequence technique described in detail in \citet[][see also \citealt{Papovich:2008aa}]{Muzzin:2013aa}.
Four high-redshift cluster candidates (see Table \ref{tbl-clusters}) were identified within the \textit{Spitzer} Adaptation of the Red-Sequence Cluster Survey \citep[SpARCS;][]{Wilson:2009ws,Muzzin:2009jm} using a two-color cut on \textit{Spitzer} IRAC 3.6~$\mathrm{\mu}$m  - 4.5~$\mathrm{\mu}$m color and $z^\prime$ - 3.6 $\mathrm{\mu}$m color.
Spectroscopic follow-up was performed using the MOSFIRE \citep{McLean:2010aa,McLean:2012aa} spectrograph on the Keck Telescopes and the Focal Reduction and Imaging Spectrograph 2 (FORS2, \citealt{Appenzeller:1998aa}) on the European Southern Observatory (ESO) Very Large Telescope (VLT).
Spectra were also obtained from the OzDES survey \citep{Yuan:2015aa,Childress:2017aa}.


\subsection{Photometric Catalogs}

Spectroscopic confirmation of these clusters was followed by collecting optical imaging data in $u^\prime\ g^\prime\ r^\prime\ i^\prime$ bands.
For SpARCS-J0330, SpARCS-J0224, and SpARCS-J0335, these data were taken with IMACS at Magellan/Baade, while for SpARCS-J0225 these data come from the Canada-France-Hawaii Telescope (CFHT) Legacy Survey (CFHTLS) which used MegaCam on CFHT.
All four clusters were imaged in near-infrared \textit{Y}-, and \textit{Ks}-band with HAWK-I at VLT, with additional \textit{J}-band photometry taken for SpARCS-J0224 and SpARCS-J0330.
Our photometry also includes the IRAC data from the Spitzer Wide-area Extragalactic Survey \citep[SWIRE;][]{Lonsdale:2003ow} with additional deeper observations in IRAC 3.6 and 4.5 $\mathrm{\mu}$m bands as part of the \textit{Spitzer} Extragalactic Representative Volume Survey (SERVS), and $z^\prime$-band data from the SpARCS survey taken by the MOSAIC-II camera at the Cerro Tololo Inter-American Observatory (CTIO).

As described in detail in \citet{Nantais:2016aa}, the imaging data were combined into a PSF-matched photometric catalog by first using Source Extractor \citep{sextractor} to detect sources in the \textit{K$_s$}-band data.
Astrometric and pixel-scale matching was performed on all images using SWarp \citep{swarp} prior to photometry.
PSF matching was performed using IRAF to generate convolution kernels before matching $u^\prime~g^\prime~r^\prime~i^\prime~z^\prime$~\textit{Y~J~Ks} band data to the poorest image quality among these bands.
Aperture photometry was performed using Source Extractor in dual-image mode and was corrected for Galactic extinction using \citet{1998ApJ...500..525S} dust maps and a \citet{2011ApJ...737..103S} extinction law.
Robust photometric errors were calculated by directly measuring the 1-$\sigma$ variation in background flux in randomly-placed apertures that do not contain any sources.

The resulting catalog has photometry in $u^\prime~g^\prime~r^\prime~i^\prime~z^\prime$~\textit{Y~J~Ks} and 3.6, 4.5, 5.8, 8.0 $\mu$m.
We perform an RA/DEC matching to the FORS2 and MOSFIRE spectroscopic data to associate spectroscopic redshifts to galaxies where possible.
Altogether there are 136 spectroscopically-confirmed members across the four clusters in this sample (see Table \ref{tbl-clusters}).

\subsection{Photometric Redshifts}\label{sec-eazy}

With the publicly-available photometric redshift code EAZY \citep{Brammer:2008uk}, we fit the broadband photometry of each object in our photometric catalog to a linear combination of seven basis templates derived from the prescription in \citet{Blanton:2007ft}.
These templates have been optimized for deep optical-NIR broad-band surveys, and this code was optimized specifically for \textit{K$_s$}-selected samples such as our own.
The output of this code includes the best-fit SED, a photometric redshift, and the photometric redshift probability distribution function of the object.
When a spectroscopic redshift is available, EAZY fixes the best-fit redshift to this value.

\subsubsection{Photometric Redshift Membership Criterion}\label{sec-members}

For our analysis, we require a cluster galaxy selection that minimizes bias toward either star-forming or quiescent galaxies.
We therefore adopt the photometric cluster membership criterion that \citet{van-der-Burg:2013aa} and \citet{Nantais:2016aa,Nantais:2017aa} used previously with this data set, and consider galaxies to be cluster members if $(z_{\mathrm{phot}}-z_{\mathrm{cluster}})/(1+z_{\mathrm{cluster}}) \leq 0.05$.
This membership criterion attempts to avoid biasing our sample, while using a range in photometric redshifts that closely matches the scatter of our photometric redshifts $(\sigma \sim 0.04)$.
The choice of 0.05 does not drive the results of this work, and repeating the analysis for cutoff values between 0.05 and 0.1 does not change our conclusions.

The selection necessarily introduces some contamination by field galaxies due to uncertainty in the photometric redshift estimates.
A previous analysis by \citet{van-der-Burg:2013aa} of a comparable data set and method shows that the overall rate of false positives and negatives is small and largely insensitive to galaxy type at $z\sim1$, indicating that this selection minimizes any error introduced to our conclusions.

\subsection{Rest-Frame Colors and UVJ Classification}\label{sec-UVJ}

To start, we perform a preliminary classification of star-forming and quiescent galaxies using the rest-frame \textit{UVJ} method \citep{Wuyts:2007aa,Williams:2009tt,Whitaker:2011aa,Patel:2012ab,van-der-Burg:2013aa,Whitaker:2013rz,muzzin2013,Strazzullo:2013aa}.
First we infer rest-frame absolute magnitudes for each cluster member by convolving its best-fit SED (derived using EAZY) with filter curves at the redshift of each galaxy.
We note that the span of the observed filters ensures that rest-frame magnitudes are interpolated from the available data, often overlapping with multiple observed passbands.
The classification is accomplished by dividing the space of rest-frame \textit{U-V} and \textit{V-J} colors into a star-forming and a quiescent region.
The cuts we use to define these regions have been empirically calibrated by \citet{Williams:2009tt} to maximally reflect the bimodality of galaxy populations as a function of redshift out to $z\sim2.5$.

In Figure \ref{fig-cmr}, we plot rest-frame \textit{U-V} vs \textit{M$_J$} color-magnitude diagrams for all cluster members in the sample, with inset rest-frame \textit{U-V} versus \textit{V-J} color-color diagrams.
Galaxies are colored according to their \textit{UVJ} classification, separating into a red-sequence and blue cloud.

\subsection{Stellar Masses and Dust Reddening}\label{sec-fast}

Using the publicly-available SED fitting code FAST \citep{Kriek:2009eq}, we fit the 12-passband photometry of each cluster to \citet[hereafter BC03]{Bruzual:2003by} stellar population synthesis templates.
FAST proceeds by generating a grid of synthetic SEDs of stellar populations at the redshift of each galaxy from the given population synthesis templates, for a range of star formation histories (SFH), ages, and masses, with possible additional variation in dust attenuation and/or metallicity.
Best-fit stellar populations are then selected from this grid by minimizing ${\chi}^2$ when comparing the synthetic SED to the observed broad-band photometry of a given galaxy.

For our grid of parameters, we use a range of ages from 100 Myr to 10 Gyr (excluding ages greater than the age of the universe at the observed redshift) and an A$_\mathrm{V}$ ranging from 0 to 3 mag with a Calzetti extinction law \citep{Calzetti:2001hh}.
Throughout, we assume an exponentially-declining star formation history, along with a Chabrier IMF \citep{chabrierimf} and fixed (solar) metallicity of 0.02.

\begin{figure*}
\centering \includegraphics[width=\textwidth]{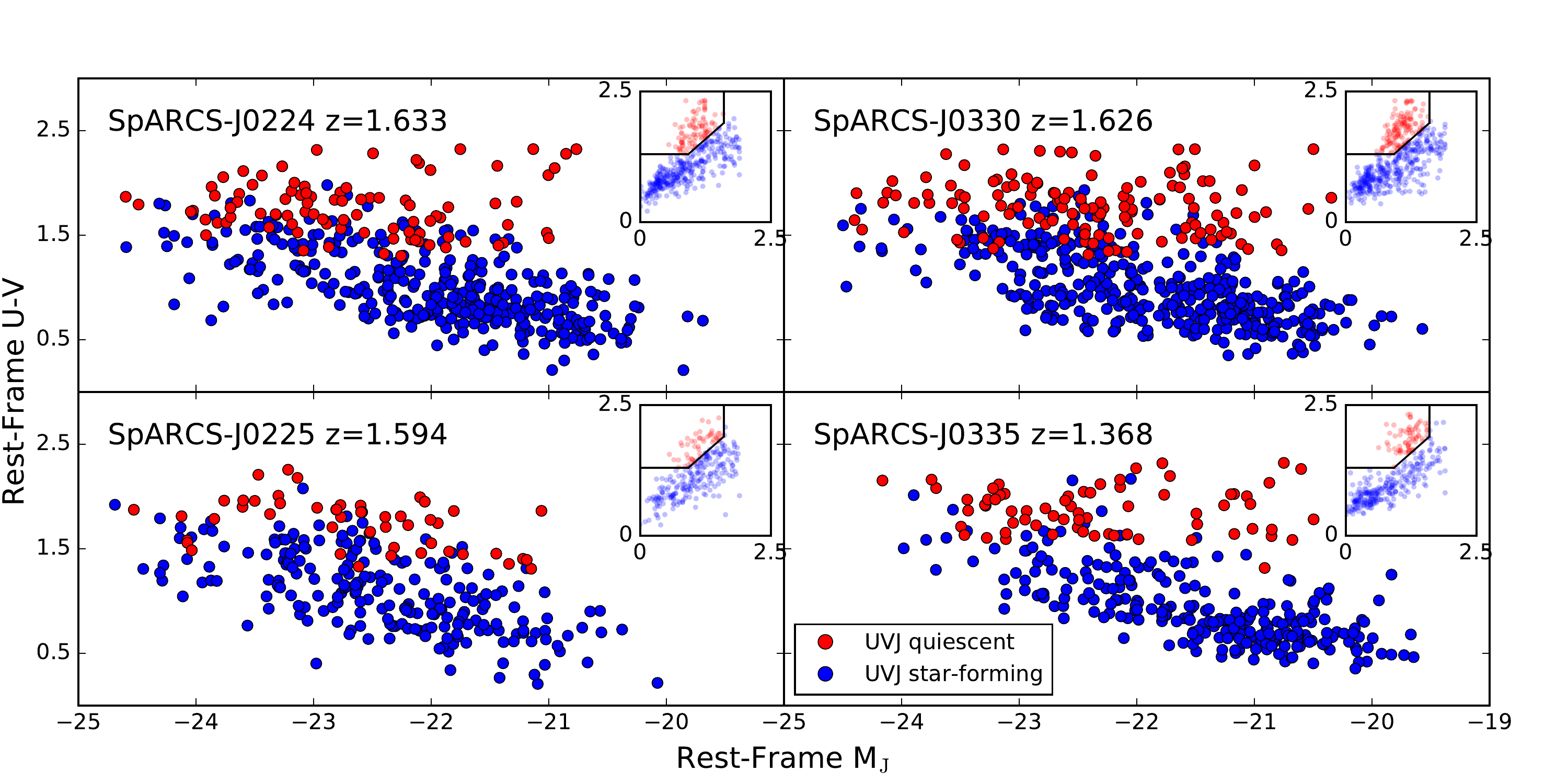}
\caption{Rest-frame \textit{U}-\textit{V} versus absolute \textit{J} magnitude $(\mathrm{M_J})$ diagram for all photometric-redshift-selected cluster members of the four clusters in the sample (see Table \ref{tbl-clusters}).
The inset panels show rest-frame \textit{U-V} versus \textit{V-J} color-color diagrams, and galaxies are colored red (quiescent) or blue (star-forming) according to their \textit{U-V} and \textit{V-J} colors (see Section \ref{sec-UVJ}).
The mass completeness of our sample corresponds roughly to a magnitude limit of $M_J \lesssim -23$.
\label{fig-cmr}}
\includegraphics[width=\textwidth]{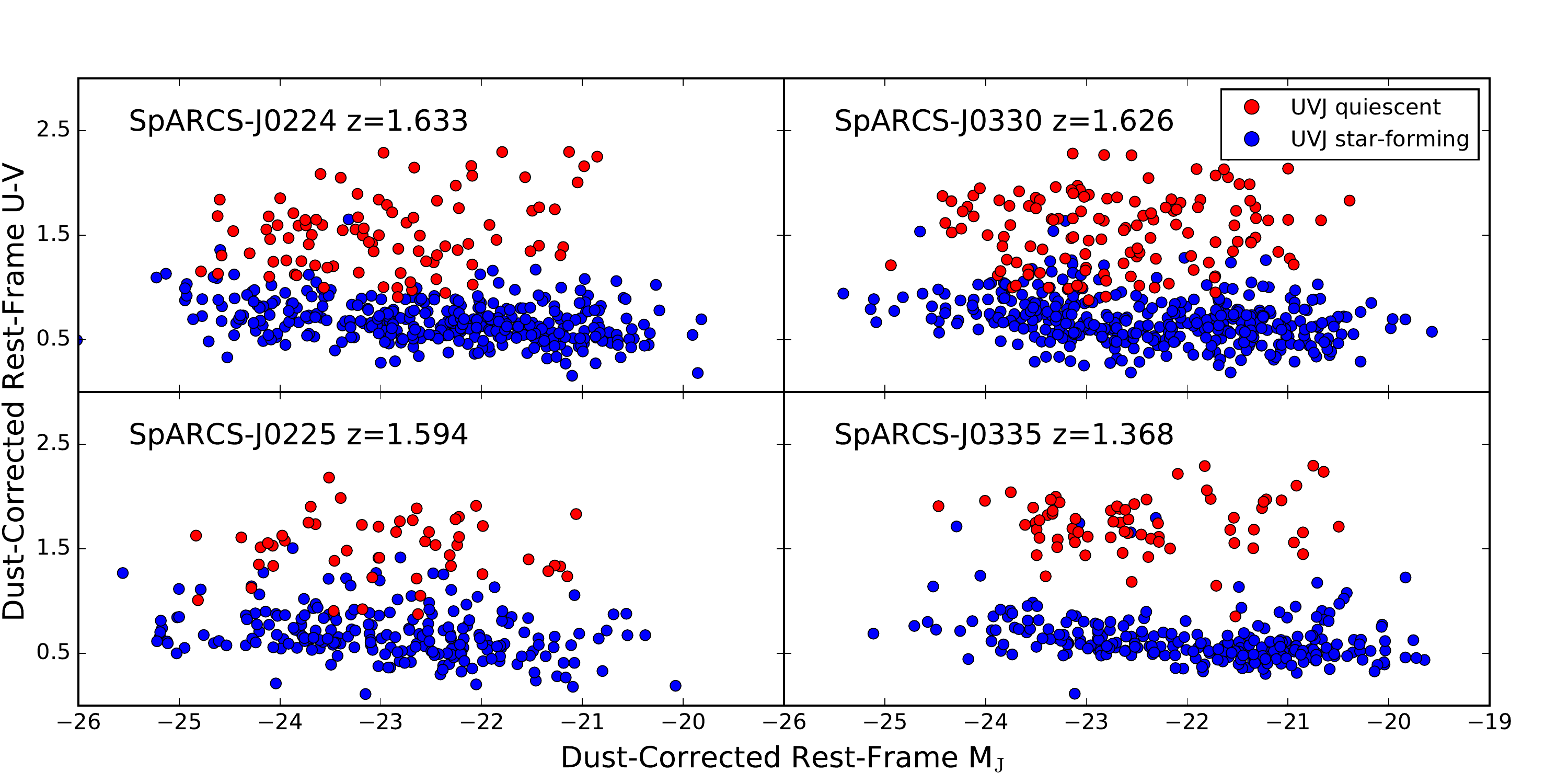}
\caption{Dust-corrected rest-frame \textit{U-V} versus absolute \textit{J} magnitude $(\mathrm{M_J})$ diagram for the four clusters in our sample.
Galaxies are colored as in Figure \ref{fig-cmr}.
Photometry is corrected for dust using a Calzetti \citep{Calzetti:2001hh} extinction law with A$_\mathrm{V}$ determined from SED fitting (see Section \ref{sec-fast}).
Compared to Figure \ref{fig-cmr}, the blue cloud reaches brighter magnitudes and exhibits smaller scatter in \textit{U-V} color.
The separation between the \textit{UVJ}-star-forming and \textit{UVJ}-quiescent populations is more apparent following dust subtraction.
\label{fig-cmr-dustsub}}
\end{figure*}

In the \textit{U-V} versus \textit{M$_J$} color-magnitude diagram of Figure \ref{fig-cmr}, galaxies segregate into a blue cloud and red-sequence.
The colors of these two populations reflect the underlying bimodal distribution in star formation rate, but this picture is complicated by the presence of star-forming galaxies with dust-reddened colors.
We therefore find it illustrative to plot the dust-corrected \textit{U-V} versus \textit{M$_J$} color-magnitude diagram in Figure \ref{fig-cmr-dustsub}.
To correct the photometry for dust, we first calculate the dust extinction in \textit{U} and \textit{V} bands for each galaxy from the total \textit{V}-band extinction (A$_\mathrm{V}$, determined through SED fitting), using a Calzetti extinction law \citep{Calzetti:2001hh}.
We then subtract the contribution from dust from each galaxy's rest-frame \textit{U} and \textit{V} magnitudes to derive the dust-corrected values of these magnitudes and colors.

Comparing Figures \ref{fig-cmr} and \ref{fig-cmr-dustsub}, we note that the red-sequence is mostly unaffected by dust subtraction, as the quiescent population generally exhibits little dust reddening to begin with.
The blue cloud becomes brighter, with dust corrections between $0-2$ magnitudes, and spans a wider range in $M_J$, while exhibiting decreased scatter in \textit{U-V} color.
The \textit{UVJ}-star-forming and \textit{UVJ}-quiescent populations separate more cleanly in color-magnitude space following dust subtraction, exposing the intermediate green valley.

\section{Analysis}\label{sec-analysis}

In this section we describe the method used to measure the quenching timescale $t_Q$.
In Section \ref{sec-model}, a toy model relates the number of star-forming, intermediate, and quiescent cluster members to a quenching timescale.
In Section \ref{sec-ellipses} we describe cluster member classification and counts.
In Section \ref{sec-r} we describe our clustercentric radial cut, and a background subtraction is described in Section \ref{sec-bkg}.
Section \ref{sec-error} describes how we derive confidence intervals for $t_Q$ with a Monte Carlo method.

\subsection{Quenching Timescale Model and Mass Completeness Limit}\label{sec-model}

A galaxy that is actively forming stars will have blue optical colors dominated by the bright contributions of short-lived O- and B-class stars.
After the onset of quenching, a galaxy's colors will become redder as these high-mass stars exhaust their hydrogen fuel and leave the main sequence, without new stars to replace them.
Eventually, a quiescent galaxy's color will reflect primarily the red colors of low-mass, long-lived main sequence stars and red giants.
We define the quenching timescale as the time since first infall after which galaxies are quenched.
In this section, we provide a conceptual summary of the method we use to measure $t_Q$, and refer the reader to Appendix \ref{sec-math} for details.

\begin{figure}[h!]
\centering \includegraphics[width=0.5\textwidth]{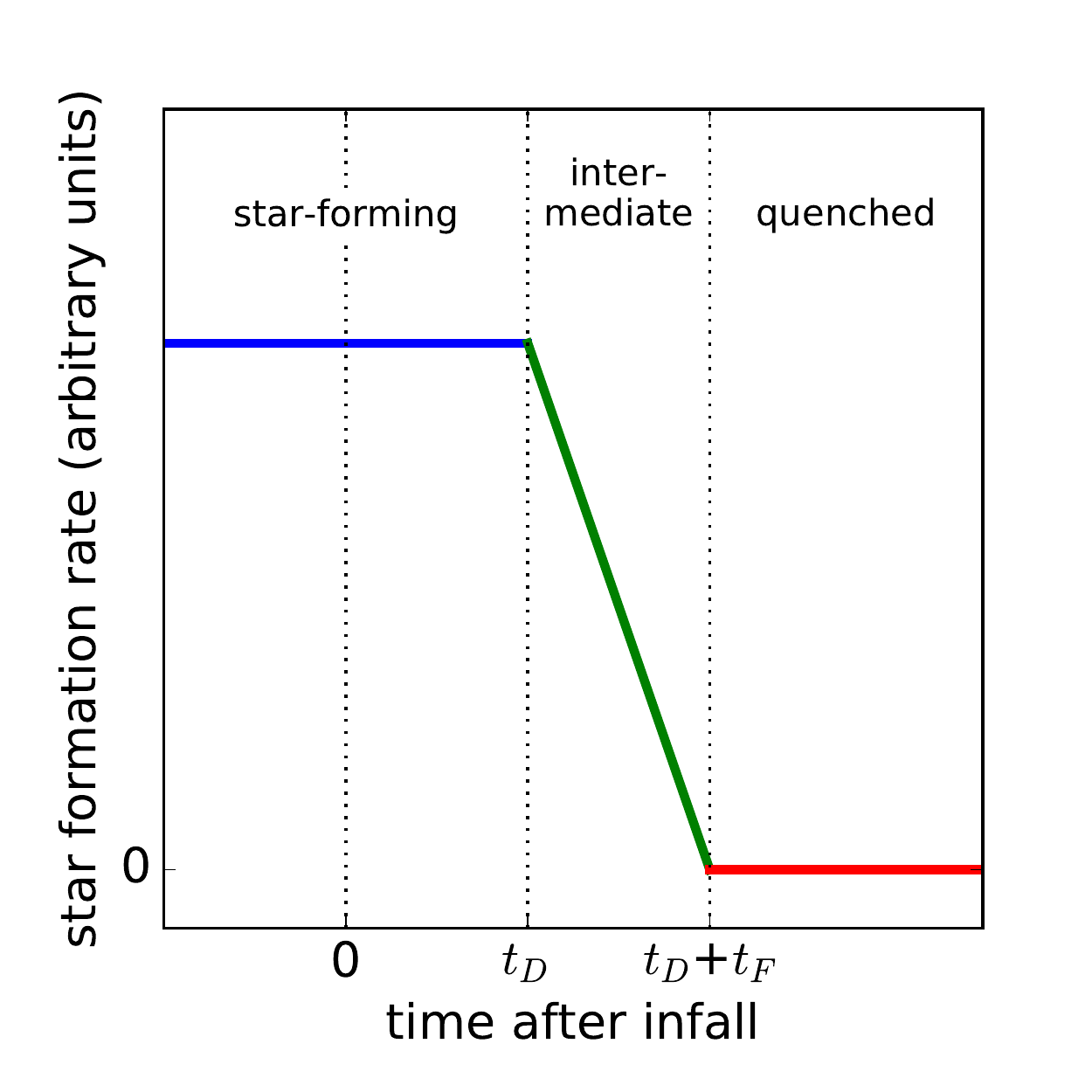}
\caption{Model of galaxy star formation rate as a function of time since infall. In this model, galaxies are star-forming and blue before being accreted by a cluster. They remain blue for a time $t_D$, the delay time, before they start to quench and become green. After a further time $t_F$, the fade time, star formation has ceased and the galaxy becomes red.
\label{fig-sfr}}
\end{figure}

Recent work has shown that environmental quenching can be described by two principal timescales, a ``delay time" ($t_D$) and a ``fade time" ($t_F$) \citep{Wetzel:2013aa, McGee:2014aa, Mok:2014aa, Haines:2015aa, Balogh:2016aa, Fossati:2017aa}.
In our model, a star-forming (blue) galaxy that is accreted by the cluster will remain blue for a time $t_D$ following infall, after which the onset of quenching causes it to become an intermediate (green) galaxy.
The galaxy will remain green for a time $t_F$, until star formation has ceased and it is quiescent (red).
This model of environmental quenching is shown schematically in Figure \ref{fig-sfr}.
The total quenching time $t_Q$, defined as the length of time after accretion until a galaxy is completely quenched, is then $t_D + t_F$.

We assume that infalling galaxies are accreted from the field.
Not every galaxy accreted from the field will be star-forming, especially at higher stellar masses, and lower redshifts.
We wish to eliminate from consideration those galaxies that were quenched in the field before they were accreted by the cluster.
We account for this by removing a fraction of quiescent galaxies proportional to the field quiescent fraction.
This fraction, as a function of redshift, can be calculated from the COSMOS/UltraVISTA \citep{McCracken:2012aa} field galaxy mass functions computed by \citet{Muzzin:2013ab}.
After subtracting the number of galaxies that were quenched at the time of infall, we can assume that the remaining galaxies were star-forming at the time of infall.
This field-quenched correction is described in detail in Appendix \ref{sec-a-field}, and for the remaining discussion we will assume corrected values.

Following the above considerations, the observed number counts of blue and green galaxies in a cluster are proportional to the length of time a galaxy spends in the delay and fade phases.
For example, a long fade time would make it easier to catch galaxies in the process of quenching, leading to larger observed numbers of green galaxies in a cluster.
To quantify these timescales in an absolute sense, one needs to control for the galaxy accretion rate of a cluster, as a higher accretion rate leads to larger numbers of all types of galaxies.
With the added assumption of a cluster galaxy infall rate, the number counts of red, green, and blue galaxies can constrain the timescales $t_D$ and $t_F$.

Given that blue galaxies have not resided in the cluster any longer than one $t_D$, their number will be equal to the cluster galaxy accretion rate $\mathrm{d}N/\mathrm{d}t$ integrated between the time of observation and one $t_D$ earlier.
In a similar manner, the number of green galaxies will be equal to the galaxy accretion rate integrated between one $t_D$ and one $t_D+t_F$ earlier.
The red galaxies trace all mass accreted earlier than one $t_D+t_F$ ago.
We write

\begin{empheq}{align*}
&B = \int_{-t_D}^{0} \mathrm{d}N/\mathrm{d}t\ \mathrm{d}t\\
&G = \int_{-(t_D+t_F)}^{-t_D} \mathrm{d}N/\mathrm{d}t\ \mathrm{d}t\\
&R = \int_{-t_H}^{-(t_D+t_F)} \mathrm{d}N /\mathrm{d}t\ \mathrm{d}t
\end{empheq}

where R,G, and B are the number of red, green, and blue galaxies respectively, $t_H$ is the Hubble time, and negative signs indicate that these galaxies were accreted in the past.

We assume that the cluster galaxy accretion rate $\mathrm{d}N/\mathrm{d}t$ is proportional to the cluster halo mass accretion rate $\mathrm{d}M/\mathrm{d}t$ as derived from the Millennium-\textsc{II} simulation by \citet{Fakhouri:2010aa}.
From there, ratios of the observed numbers of R, G, and B galaxies can be related to $\mathrm{d}M/\mathrm{d}t$, $t_F$, and $t_D$, to constrain the fade and delay times and thereby the total quenching time.
In Appendix \ref{sec-math} we more fully describe this toy model, which is ultimately defined by a set of four equations, \eqref{eq-model-first}~--~\eqref{eq-model-last}.
Given a number of R, G, and B galaxies, a cluster redshift, and a mass accretion rate, Equations \eqref{eq-model-first}~--~\eqref{eq-model-last} can be solved for $t_F$, $t_D$, and $t_Q$.

Before proceeding with the analysis, we note several considerations which must be taken into account with this model.
The $80\%$ mass completeness of our sample is defined as the lowest mass for which passive galaxies yield accurate passive fractions \citep{van-der-Burg:2013aa}.
This limit varies from $10^{10.3}$ to $10^{10.5}\ \mathrm{M}_\odot$ within our sample \citep{van-der-Burg:2013aa,Nantais:2016aa}, due to variations in exposure time and redshift.
We must restrict our analysis to galaxies with masses above these limits, to ensure a fair comparison between the quenched and not-yet-quenched galaxies.

Second, it has been shown that the environmental quenching timescale varies with satellite galaxy mass \citep{De-Lucia:2012aa,Wetzel:2013aa,Wheeler:2014aa,Fillingham:2015aa}, and it is therefore inaccurate to refer to a singular environmental quenching timescale for all galaxies.
Any quenching timescale measured with the above toy model will necessarily be for an ensemble of galaxies spanning some range in stellar mass.
However, the quenching timescale does not vary much over the small dynamical range in mass studied in this work, at least at low redshift \citep[e.g. see Fig. 8 of ][Figure 5]{Fillingham:2015aa,Wetzel:2013aa}.

Third, the mass dependence must be considered when comparing with results of different studies.
Comparing with other studies will allow us to investigate the evolution of $t_Q$ with redshift (see Section \ref{sec-discussion}).
Other measurements of $t_Q$ will not be comparable to our results unless they were derived for a similar mass range.

For the above reasons, when measuring $t_Q$ we restrict our sample to galaxies with stellar masses above a mass completeness limit $M_* \ge 10^{10.5}~ \mathrm{M}_\odot$.
This cut conservatively ensures that we are sampling above the mass completeness of our photometry for each cluster, and allows comparison with various results in the literature that report the quenching timescale for this range of masses.

In general, environmental quenching is likely the result of several different mechanisms operating over different timescales and environments \citep{Schawinski:2014aa,Paccagnella:2016aa,Paccagnella:2017aa}.
A toy model such as the one presented here is not intended to be a final description of environmental quenching, but instead to investigate which physical scenarios, if any, are consistent with a set of very simple assumptions.

\subsection{Classification of Galaxies as Star-Forming, Intermediate, or Quiescent}\label{sec-ellipses}

The environmental quenching model described in Section \ref{sec-model} and Appendix \ref{sec-math} relates the number of observed star-forming (blue), intermediate (green), and quiescent (red) cluster members to the delay and fade times, $t_D$ and $t_F$.
A method of classifying galaxies as red, green, or blue is therefore needed before we can solve for the quenching timescale, $t_Q$.
We will describe a new classification method, not to be confused with the preliminary \textit{UVJ}-quiescent and -star-forming classification performed in Section \ref{sec-UVJ}, as the \textit{UVJ} method lacks an intermediate (green) category (see Section \ref{sec-uvj-problem}).

A common approach to identifying star-forming, intermediate, and quiescent galaxies is to categorize them according to their colors and magnitudes, in a manner informed by galaxy evolutionary models.
A successful classification scheme will distinguish between star-forming galaxies that appear red due to dust, and galaxies that are red from a lack of star formation.
In this section we introduce a classification based on dust-corrected rest-frame colors derived from SED fitting (see Sections \ref{sec-UVJ} and \ref{sec-fast}).

Each galaxy's best-fit SED parameters include the \textit{V}-band dust reddening $\mathrm{A_V}$, which we use in conjunction with a Calzetti extinction law \citep{Calzetti:2001hh} to determine the reddening in \textit{U}- and \textit{B}-bands.
Subtracting this reddening from the rest-frame photometry breaks the color degeneracy between dusty, star-forming galaxies and old, quiescent galaxies.
Following dust-subtraction, galaxies separate more cleanly into a red-sequence, green valley, and blue cloud in a color-magnitude diagram, such as those shown in Figure \ref{fig-cmr-dustsub}.
We can therefore use cuts in dust-corrected color-magnitude space to label galaxies red, green, or blue.

\begin{figure}[h!]
\centering \includegraphics[width=0.5\textwidth]{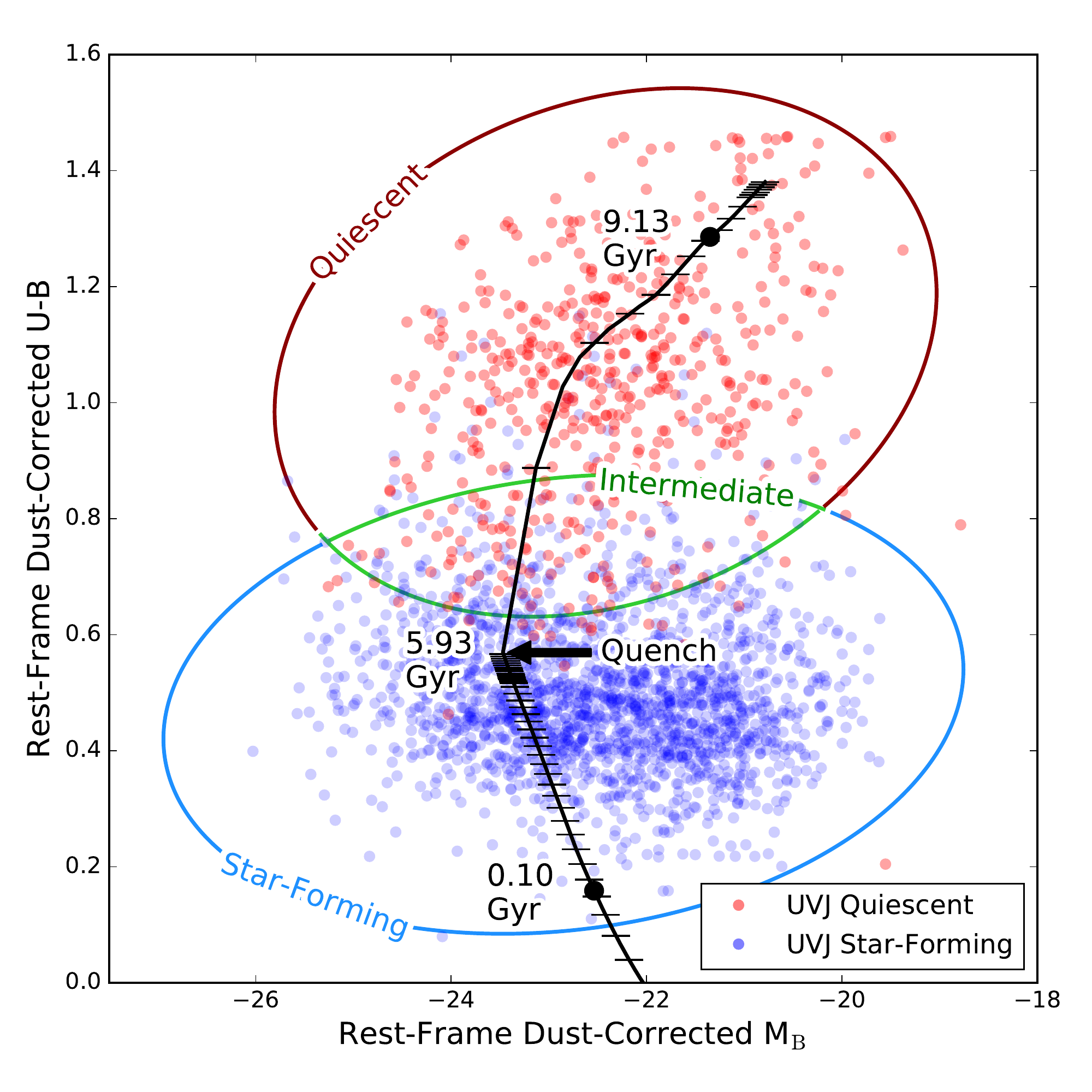}
\caption{Classification of star-forming, intermediate, and quiescent galaxies.
We plot the dust-corrected rest-frame \textit{U-B} versus absolute \textit{B} magnitude for all cluster members.
Points are colored according to galaxies' \textit{UVJ} classifications (see Section \ref{sec-UVJ}).
The colored lines show $3\mhyphen\sigma$ elliptical fits to the two principal clusters of data points identified by a spectral clustering algorithm.
The elliptical regions define the quiescent, intermediate, and star-forming populations of galaxies, as labeled.
The solid black line is a BC03 model evolutionary track for continual star formation that truncates after 6 Gyr.
The black line is punctuated by dashes indicating time intervals evenly spaced in redshift.
The black points on this line mark when the model is is $0.10$ and $9.13$ Gyr old.
This track demonstrates good agreement between the model's star formation rate and its progressive classification from blue, to green, to red.
Note that even after 6 Gyr of constantly-integrated star formation, the model remains fully within the star-forming ellipse, only leaving it after quenching.
\label{fig-ellipses}}
\end{figure}

To define these cuts, we start by applying a spectral clustering algorithm to the dust-corrected color-magnitude diagram of all galaxy cluster members.
This algorithm labels the two principal clusters of data points, identified in this case with the blue cloud and red-sequence.
We then fit an elliptical region to each cluster of data points by finding the eigenvectors of the covariance matrix of the set of points, which define the semi-major and semi-minor axes of an ellipse.
The width and height of this ellipse are scaled so that the ellipse represents a 95\% ($\operatorname{2-\sigma}$) confidence level.\footnote{Specifically, the length of each elliptical axis is $4\sqrt{\lambda}$, where $\lambda$ is the eigenvalue of the axis's eigenvector.}
Galaxies are categorized as either star-forming or quiescent according to their membership in these elliptical regions.
We define the green valley as the overlapping area of these ellipses, and galaxies within this region are categorized as intermediate.
In Figure \ref{fig-ellipses} we plot the classification regions over the dust-corrected rest-frame colors and magnitudes of all cluster members.

For comparison, we include on this plot a BC03 evolutionary track for a stellar population with a star formation rate that remains constant for 6 Gyr, after which it truncates (quenches).
There is a clear agreement between the model's stage of evolution and its progressive classification from blue, to green, to red.
In its star-forming phase, a galaxy stays in the blue region, and doesn't enter the green (intermediate) region until it is quenched.
After quenching, the model crosses the green valley in $\sim0.2$ Gyr.
The straightforward nature of galaxy evolution in this dust-subtracted color-magnitude space is the primary advantage of this classification scheme, which identifies an unambiguous green valley between the blue cloud and red-sequence.

These elliptical regions define the star-forming, quiescent, and intermediate populations, and therefore the final value of $t_Q$ depends on their precise contours.
The total value of $t_Q$ is set by the location of the border between the green and red population, while the blue-green border, determining the fraction of star-forming galaxies that are intermediate, affects the way $t_Q$ is subdivided into $t_D$ and $t_F$.
Through repeated experimentation, we determine that reasonable tweaks to the contours of these ellipses affect the resulting $t_Q$ within error bars.
The red-green border necessarily lies in the green valley, a region of low galaxy number density.
The total quenching time is therefore robust to small adjustments in this border, as the bulk of galaxies that are considered quenched or star-forming are not affected.

\subsubsection{Alternative UVJ-based Classification}\label{sec-uvj-problem}

We also investigate the impact of alternatively using rest-frame \textit{U-V} versus \textit{V-J} color-color diagrams to classify star-forming and quiescent galaxies (see Section \ref{sec-UVJ}).

The location of the green valley in \textit{UVJ} space is not clear.
Accordingly, the quenching timescale model must be simplified to forgo the use of green galaxies.
This simplification comes at the cost of being unable to constrain separate delay and fade times $t_D$ and $t_F$, instead directly measuring the total quenching time, $t_Q$.

For details on the results of this approach, we refer the reader to Appendix \ref{sec-a-uvj}.
This subject will be further elaborated in a letter (Foltz 2018, in prep).

\subsection{Projected Radial Distance Cut}\label{sec-r}

A cluster galaxy's type and time since infall correlates with clustercentric distance.
We wish to compare and combine galaxy number counts across multiple clusters and cluster samples, and therefore must control for galaxies' locations within the cluster.
Although a cut based on galaxies' positions relative to the cluster's virial radius is commonly used for this purpose, it is unlikely that the clusters in the high-redshift cluster sample are completely virialized structures.
Because of this, it would not be meaningful to naively ascribe virial radii to the velocity dispersions that we measure.

We therefore test our method using a variety of cuts on physical clustercentric distance, $r \leq 1000$ kpc, $r \leq 1500$ kpc, and $r \leq 2000$ kpc.
The choice of radial cut does not greatly affect the results of our analysis, and so we choose to restrict our consideration to galaxies with $r \leq 2000$ kpc.

\subsection{Background Subtraction}\label{sec-bkg}

Our number counts are contaminated by the inclusion of field galaxies due to inherent uncertainty in our photometric-redshift selection.
Before determining the quenching timescale we need to subtract the field galaxy background.
We therefore adjust the number counts for each cluster to correct for field contamination estimated from the field galaxy survey catalogs from UltraVISTA/COSMOS \citep{muzzin2013}.

To estimate the number of field galaxies included in the cluster sample, we start by cropping a randomly-selected section of the Ultra-VISTA/COSMOS dataset to match the angular size of the cluster photometry.
We process the Ultra-VISTA/COSMOS photometry with EAZY and FAST (see Sections \ref{sec-eazy} and \ref{sec-fast}) to determine photometric redshifts, rest-frame colors, and masses, limiting the data set to the same photometric bands that are available in the main dataset.
We then select field galaxies from this sample at the redshift of the cluster based on the same photometric redshift criterion described in Section \ref{sec-members}.
These field galaxies are classified as star-forming, intermediate, or quenched, according to the dust-corrected color-magnitude cuts detailed in Section \ref{sec-ellipses}.
We then subtract these numbers of red, green, and blue field galaxies from the corresponding numbers of cluster galaxies.

\subsection{Uncertainty Calculation}\label{sec-error}

Shown in detail in Appendix \ref{sec-math}, the numbers of red, green, and blue cluster galaxies, together with a cluster redshift, fully determine a quenching timescale.
The uncertainty in $t_Q$ is driven by uncertainty in these number counts, and we therefore use a Monte Carlo method to estimate the $68\%$ confidence interval for $t_Q$.

For each cluster, we create 200 simulated data sets by varying the rest-frame photometry of each galaxy by a random amount drawn from a normal distribution defined by the galaxy's rest-frame photometric error bars.
For each simulated data set we then count the numbers of R, G, and B galaxies and substitute these counts into Equations~\eqref{eq-model-first}~--~\eqref{eq-model-last} and solve for $t_Q$, arriving at a distribution in $t_Q$.
The central $68\%$ of this distribution then defines the upper and lower confidence intervals for $t_Q$.

\section{Results}\label{sec-results}

\begin{deluxetable*}{ccccccccc}
\tabletypesize{\scriptsize}
\tablecaption{Quenching timescale measured in the SpARCS and GCLASS cluster samples\label{tbl-results}}
\tablecolumns{9}
\tablewidth{0pt}
\tablehead{
\colhead{Cluster} \vspace{-0.4cm} & & & & & & \colhead{$t_D$} & \colhead{$t_F$} & \colhead{$t_Q$} \\ \vspace{-0.4cm}
& \colhead{N$^\mathrm{a}$} & \colhead{$\bar{z}$} & \colhead{R$^\mathrm{b}$} & \colhead{G$^\mathrm{b}$} & \colhead{B$^\mathrm{b}$} & & & \\
\colhead{Sample} & \colhead{} & \colhead{} & \colhead{} & \colhead{} & \colhead{} & \colhead{(Gyr)} & \colhead{(Gyr)} & \colhead{(Gyr)}
}
\startdata
GCLASS & 10 & 1.04 & 160 & 42 & 38 & \gtd & \gtf & \gresult\\
SpARCS high-redshift & 4 & 1.55 & 79 & 17 & 63 & \htd & \htf & \hiresult\\
\enddata
\tablenotetext{a}{Number of galaxy clusters in the sample.}
\tablenotetext{b}{Number of red, green, or blue galaxies above the mass completeness limit.}
\end{deluxetable*}

Here we report the results of the quenching timescale modeling described in Section \ref{sec-model}.
In Section \ref{sec-z1.6} we report the measured quenching timescale for our high-redshift sample.
In Section \ref{sec-z1} we report the quenching timescale measured in a sample of galaxy clusters at $z\sim1$, and compare with a previous, independent measurement of the same reported by \citet{Muzzin:2014aa}.

The results are summarized in Table \ref{tbl-results}.

\subsection{Quenching timescale at $z=1.55$}\label{sec-z1.6}

We start by selecting cluster members according to the photometric redshift probability cut defined in Section \ref{sec-members}.
We classify galaxies as red, green, or blue according to their colors and magnitudes by the method described in Section \ref{sec-ellipses}.
We stack the sample by taking the total number of red, green, and blue galaxies at the mean redshift of the cluster sample, $z_c=1.55$.
We substitute these values for R, G, B, and $z_c$ into Equations~\eqref{eq-model-first}~--~\eqref{eq-model-last} and solve for $t_Q$, finding a quenching timescale of $t_Q=$\hiresult Gyr for this sample.

\subsection{Quenching timescale at $z=1.0$}\label{sec-z1}

The Gemini Cluster Astrophysics Spectroscopic Survey \citep[GCLASS,][]{Muzzin:2012dw} is a sample of 10 red-sequence-selected clusters at $0.87 < z < 1.34$, initially detected by the SpARCS optical/IR cluster survey using the cluster red-sequence detection method developed by \cite{Gladders:2000rq} \citep[see][]{Muzzin:2009jm,Wilson:2009ws,Demarco:2010om}.
GCLASS forms a complimentary data set to the $z\sim1.6$ SpARCS sample, having a similar range of optical to far-infrared photometry and catalogs prepared in a homogeneous manner \citep[see][]{Muzzin:2012dw,van-der-Burg:2013aa,Nantais:2016aa,Nantais:2017aa}.
With this data set, we can compare quenching timescales at $z\sim1.6$ and $z\sim1$.

Using the GCLASS spectroscopic and photometric catalogs, we performed the same cluster member selection and categorization described in Sections \ref{sec-members} and \ref{sec-ellipses}.
We then use the total number of red, blue, and green galaxies above the mass completeness limit $M_* \geq 10^{10.5}~\mathrm{M}_\odot$ to measure a quenching timescale according to Equations~\eqref{eq-model-first}~--~\eqref{eq-model-last}, finding $t_Q=$\gresult Gyr at $z\sim1$.

A previous analysis by our team has independently measured the quenching timescale in this sample.
\citet{Muzzin:2014aa} identified spectroscopic cluster members with absorption line features indicative of recent, rapidly-truncated star formation.
The distribution of these ``post-starburst" galaxies in phase space, when compared with the phase space of zoom simulations, indicated a quenching timescale of $\sim1\ \rpm\ 0.25$ Gyr.
This result is largely independent of the measurement performed in this present work, as it was derived using galaxies' spectroscopic features and positions within the cluster.
The agreement between these methods is therefore a strong indicator that they independently measure the same timescale, corresponding to the quenching time.

\section{Discussion}\label{sec-discussion}

\begin{figure*}[h!]
\centering \includegraphics[width=\textwidth]{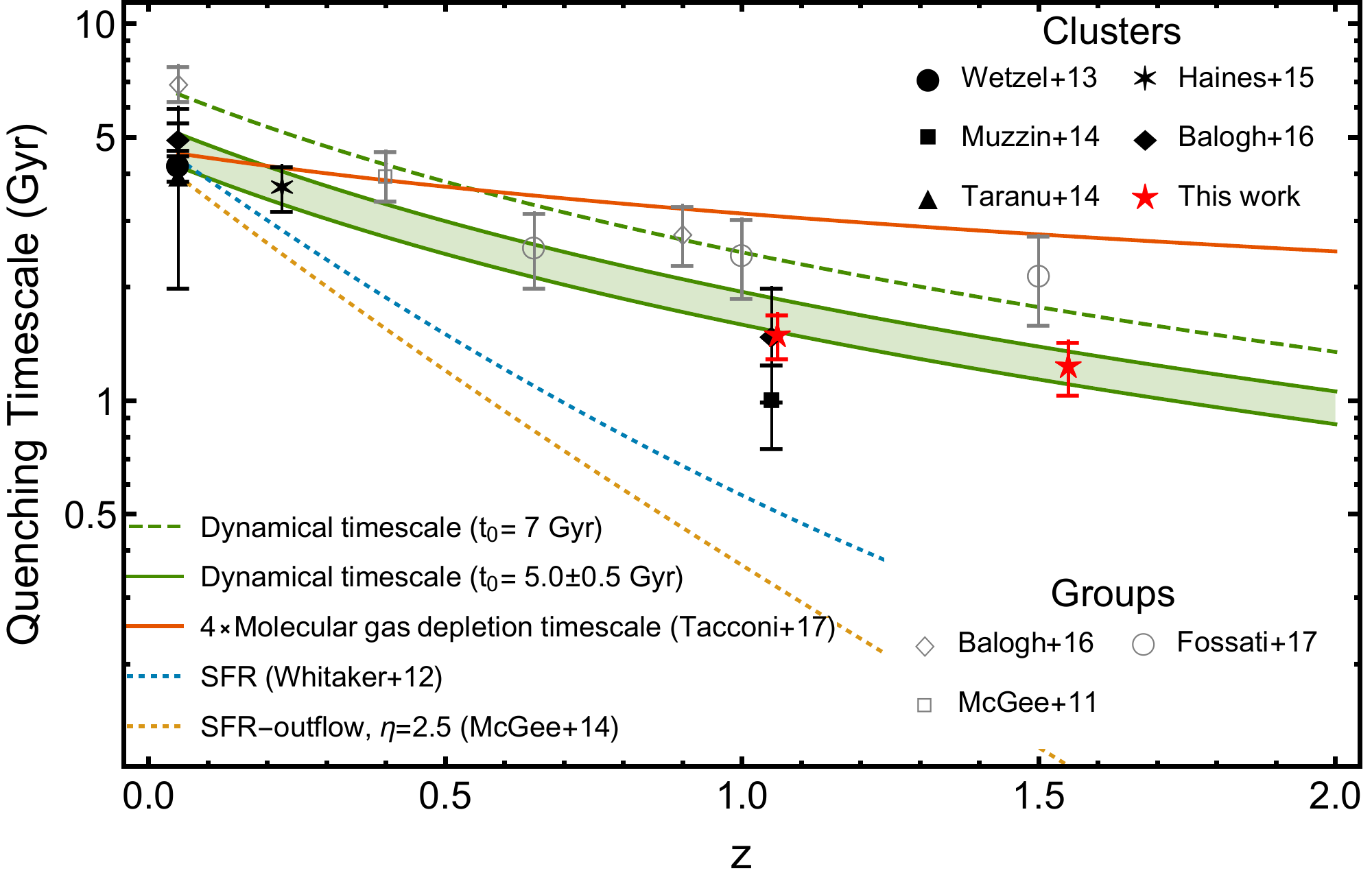}
\caption{
Quenching timescale as a function of redshift.
Red points show the quenching timescales measured for our cluster samples at $z\sim1$ and $\sim1.6$ (see Section \ref{sec-results}).
Black points show the quenching timescales measured in clusters by \citet{Wetzel:2013aa}, \cite{Muzzin:2014aa}, \citet{Taranu:2014aa}, \citet{Haines:2015aa}, and \citet{Balogh:2016aa}.
Hollow gray points indicate quenching timescales measured in groups by \citet{McGee:2011aa}, \citet{Balogh:2016aa}, and \citet{Fossati:2017aa}.
All data points are for galaxies with $M_* \ge 10^{10.5}~ \mathrm{M}_\odot$.
The dashed green line represents the evolution of a dynamical timescale normalized to 7 Gyr at $z=0.05$, the quenching time in SDSS groups as reported by \citet{Balogh:2016aa}.
The shaded green region represents the evolution of the dynamical timescale normalized to $5.0\pm0.5$ Gyr, spanning the range of quenching times in SDSS clusters as reported by \citet{Wetzel:2013aa} and \citet{Balogh:2016aa}.
The solid red line indicates a rough approximation of the total gas depletion timescale, $t_{\mathrm{depl}}(\mathrm{H_I}+\mathrm{H_{mol}})$, adapted from the molecular gas depletion timescale measured by \citet[][see text]{Tacconi:2017aa}.
The dotted blue and orange lines are estimates of the quenching time in an SFR outflow scenario.
The blue dotted line is an estimate of the gas depletion timescale with a mass loading factor of $\eta=2.5$, described by \citet{McGee:2014aa}.
The orange dotted line approximates the evolution of an outflow gas depletion time, being inversely proportional to the evolution in star formation rates of the fundamental plane as measured by \citet{Whitaker:2012aa}, normalized to the low-redshift time of 5 Gyr.
\label{fig-tq}}
\end{figure*}

Based on the results of Section \ref{sec-results}, the quenching timescale for massive satellite galaxies ($M_* \ge 10^{10.5}~ \mathrm{M}_\odot$), measured in a homogeneous manner across cluster samples, is $\simgresult$ Gyr at $z\sim1$ and $\simhiresult$ Gyr at $z\sim1.6$.
These quenching times are required to produce the observed number of quenched galaxies in our cluster sample, given a reasonable model of the mass accretion histories of clusters.
We plot the evolution of the cluster quenching timescale with redshift in Figure \ref{fig-tq}.

\subsection{Redshift Evolution of Observed Quenching Timescales}

Included on Figure \ref{fig-tq} are several quenching timescales drawn from other studies.
We note several possible sources of confusion that must be accounted for when drawing fair comparisons between timescales reported in the literature.
Historically, researchers have used several different approaches to modeling or measuring the quenching timescale, and occasionally even different definitions of the quenching timescale itself.
We have taken $t_Q$ to be the time following infall for a galaxy to be classified quiescent, and following \citet{Wetzel:2013aa}, describe it with a ``delay" followed by a ``fade" phase.
Other formalisms have been adopted, such as ``slow quenching" scenarios where galaxies begin quenching immediately upon infall, having star-formation rates that decline gradually with an exponential time constant (often also called the ``quenching time").

These considerations are additional to the normal systematic differences in galaxy samples and completenesses, classification systems, membership selections, and background subtractions.
In the end, all models must necessarily employ various simplifying assumptions, and are approximations to a full description of galaxy quenching.

The data points described here were all measured for group or cluster galaxies in stellar mass ranges equal or comparable to our mass completeness limit, $M_*~\ge 10^{10.5}~ \mathrm{M}_\odot$.
In works where the quenching timescale was reported for separate redshift or mass bins, we take the mean quenching timescale for galaxies above our mass limit, at the mean redshift of the redshift bin.
We plot cluster measurements as solid black symbols, while group measurements are plotted as hollow gray symbols.

\citet{McGee:2011aa,McGee:2014aa} studied the passive fraction in galaxy groups taken from the Group Environment Evolution Collaboration \citep[GEEC and GEEC2,][]{Balogh:2014aa}.
\citet{McGee:2014aa} relates the group passive fraction of $\sim~0.3$ at $z=0.4$ to infall histories in semi-analytic simulations \citep{McGee:2009aa}, where 30\% of galaxies became satellites more than $4.4\pm0.6$ Gyr ago.
From this, it is concluded that the quenching time for these galaxies is 4.4 Gyr.

This basic approach was adapted by \citet{Wetzel:2013aa}, \citet{Balogh:2016aa}, and \cite{Fossati:2017aa}, and applied to galaxy groups and clusters in the Sloan Digital Sky Survey \citep[SDSS,][]{York:2000aa}, GEEC2, GCLASS, and deep-field 3D-HST/CANDELS \citep{Grogin:2011aa,Koekemoer:2011aa,Brammer:2012aa} data sets.
In SDSS clusters, \citet{Wetzel:2013aa} find a total quenching time of $4.4 \pm 0.4$ Gyr, where \citet{Balogh:2016aa} finds $5.0\pm0.5$ Gyr.
\citet{Balogh:2016aa} also finds a quenching time of $7.0 \pm 0.5$~Gyr in SDSS groups, $2.8 \pm 0.5$~Gyr in GEEC2 groups, and $1.5 \pm 0.5$~Gyr in the GCLASS cluster sample.
\citet{Fossati:2017aa} reports the quenching timescale for groups in the 3D-HST/CANDELS fields in three redshift bins spanning $0.5 < z < 1.80$, finding quenching times between 2 and 3 Gyr.

\citet{Muzzin:2014aa} employ a different method to constrain quenching timescales in the GCLASS cluster sample.
Using galaxy spectral features, they identify a population of poststarburst galaxies.
The distribution of this population in cluster phase space\footnote{``Cluster phase space" here refers to the phase space spanned by galaxies' velocities relative to the cluster and their projected clustercentric radius.} can be related to the evolving phase space distribution of infalling subhalos in dark-matter zoom simulations to determine a timescale.
\citet{Muzzin:2014aa} reports that this process indicates a rapid fade time of $t_F \simeq 0.5$ Gyr following the galaxy's first pass through $0.25\mhyphen0.5~ R_{200}$, a passage which requires a time $t_D=0.45\pm0.15$~ Gyr in the simulations, for a total quenching time of $t_Q=1.00\pm0.25$~ Gyr.

Other studies have successfully measured quenching timescales, but use different models or assumptions that complicate direct comparison with the present work.
While we define $t_Q$ to be the time after accretion required for a galaxy to be classified quiescent, it is not uncommon to find the quenching timescale defined in other ways.
In the ``slow quenching" model, star-formation rates decline gradually with an exponential time constant $\tau_Q$ starting immediately upon infall.
To convert from this framework to our present system of classification, we create BC03 model stellar populations with star formation rates that remain constant until infall, after which they decline with time constant $\tau_Q$.
We then plot the evolution of the model rest-frame color and magnitude on the classification ellipses of Figure \ref{fig-ellipses}, and take $t_Q$ to be the time required after infall before the model is considered red.

\citet{Haines:2015aa} employ a similar phase-space approach to \citet{Muzzin:2014aa}, comparing the radial density profiles of star-forming galaxies in clusters at $z\sim0.2$ to the evolving radial densities of infalling halos in clusters the Millennium-\textsc{II} simulation, at a slightly lower mass completeness limit of $2\times10^{10}~ \mathrm{M}_\odot$.
They adopt the ``slow quenching" model, and find the kinematic properties of the star-forming population to be best fit by an exponentially-declining star formation rate with time constant $\tau_Q=1.73\pm0.25$~ Gyr.
The value of $t_Q$ corresponding to this result depends on the assumed age of the galaxy at time of infall.
Cluster red-sequence galaxies at $z \lesssim 1$ have colors consistent with having been formed at $z \gtrsim 3$ \citep{Foltz:2015aa}, and models of cluster mass-accretion rates suggest that a typical halo in a cluster at $z=0.2$ was accreted at $z\sim1.1$ \citep{Fossati:2017aa}.
Therefore we construct our model with an age of 3 Gyr at infall, and find that $\tau_Q=1.73\pm0.25$~Gyr corresponds to $t_Q\simeq3.7\pm0.5$~Gyr.

\citet{Taranu:2014aa} employ a novel combination of observed galaxy bulge and disc colors, models of quenching star formation rates, and subhalo orbits drawn from cosmological N-body simulations.
They too adopt a ``slow quenching" model, and their data are best fit by an exponentially-declining star formation rate time constant of $\tau_Q=3\mhyphen3.5$~Gyr, with quenching beginning immediately upon infall.
Adopting the same conversion method as we use for \citet{Haines:2015aa}, we find this corresponds to $t_Q\simeq4\pm2$ Gyr.
We note that \citet{Taranu:2014aa} use a sample of brightest cluster (and group) galaxies, an extremal population of quenched galaxies, for which our model likely breaks down.

Other notable studies preclude comparison with the present work, due to differences in mass completeness, or differences in analysis.
\citet{Oman:2016aa} use a phase space approach to characterize the quenching timescale in SDSS clusters.
\citet{Oman:2016aa} derive orbital histories for cluster and satellite galaxies from dark-matter simulations, characterizing the probability that each galaxy becomes quiescent as a function of time, $p_q(t)$.
They report a typical delay time of $t_D=3.5\mhyphen5$~Gyr and a $p_q(t)$ that evolves with a time constant $\tau\lesssim2$~Gyr.
We do not attempt to interpret this in terms of a $t_Q$ value.

\citet{Gobat:2015aa}, studying galaxies of mass $M_* \gtrsim 10^{11}~ \mathrm{M}_\odot$ in groups in the COSMOS field at $z\sim1.8$, find evidence for a rapid fade time of $t_F\approx0.3$~Gyr, based on the properties of satellite galaxies.
In the local universe, for galaxies with masses
$M_* > 10^{9.8}~ \mathrm{M}_\odot$, \citet{Schawinski:2014aa,Paccagnella:2016aa,Paccagnella:2017aa} conclude that quenching happens by separate rapid and slow-quenching scenarios.
\citet{Paccagnella:2016aa,Paccagnella:2017aa} find that intermediate galaxies are described by a slow-quenching scenario with a total timescale of 2-5~Gyr, although fast quenching of poststarbursts produces two times as many passive galaxies.

\subsection{Remarks on Methods and Systematic Error}\label{sec-systematics}

The various techniques that have been used all share two main features in common.
First, they all must label a population of quenched galaxies, and/or a star-forming population.
This is accomplished variously by cuts on colors and/or magnitude, inferred star formation rates, or galaxy spectral features.
Second, they must relate the characteristics of the quenched or active population, or quenched fraction to timescale information.
This is universally done by comparison with numerical simulations, which can relate infall times to distributions in phase space, radial surface densities, or to mass accretion histories, as in the present work.

Besides these fundamental differences in model, the next most important source of systematic error is likely the choice of how to treat the field-quenched correction (Appendix \ref{sec-a-field} in the present work).
When characterizing the quenched population of a cluster, one needs to account for the fact that the observed quenched fraction in clusters isn't entirely the result of quenching within the cluster, because quenched galaxies are found in the field as well.
Therefore some number of quenched galaxies need to be subtracted from the observed count, in a manner informed by the field quenched fraction.
For \citet{McGee:2011aa}, \citet{Balogh:2016aa}, and \cite{Fossati:2017aa}, this is done by calculating the quenched fraction that is in excess of the field at the observed redshift of the cluster, which is sometimes referred to as the ``conversion fraction" or the ``environmental quenching efficiency" \citep{van-den-Bosch:2008aa}.
The approach used by \citet{Wetzel:2013aa} and the present work is to instead subtract off those field galaxies that were quenched at the time of accretion, not at the time of observation.

As explained in Appendix B of \citet{Balogh:2016aa}, the different approaches amount to a philosophical difference about what is being measured.
By calculating the conversion fraction, one removes not only those galaxies which were quenched at the time of accretion, but also those which would have quenched in the field by the time of observation, too.
The result is that the \citet{Wetzel:2013aa} approach measures the time taken for galaxies to quench in dense environments, while the ``conversion fraction" approach measures the timescale due purely to environmental quenching.
\citet{Balogh:2016aa} found $t_Q$ to be higher by 0.5~Gyr for SDSS clusters than previous estimates by \citet{Wetzel:2013aa}, and attributes this difference to the above difference in field subtraction methods, while noting that the true answer likely lies somewhere in between.
By $z\sim1$, $t_Q$ as measured in the GCLASS cluster sample by \citet{Balogh:2016aa} and the present work agree within error bars.

For the present work, the field correction approach of \citet{Wetzel:2013aa} is necessary.
Our model requires a direct comparison between quenched galaxies and those which have not yet been quenched, under the assumption that these populations are the same except for the time they have spent in the cluster.
In other words, the model assumes that the B, G, and R populations represent an evolutionary sequence, B $\rightarrow$ G $\rightarrow$ R.
It is possible to calculate the conversion fraction of our cluster sample (see \citealt{Nantais:2016aa,Nantais:2017aa}), arriving at the number of cluster galaxies quenched due solely to environment, but these would have to be compared to only those blue galaxies that will quench due solely to environment.
It is unclear how to correct the blue population in this way without knowing the quenching timescale in advance.
We therefore adopt the convention of subtracting only those galaxies that were already quenched at the time of accretion, and therefore measure the net change in galaxy properties since infall.

Of special interest within the assembled data set is a comparison between the three studies that have measured the quenching timescale in the GCLASS sample \citep[][and the present work]{Muzzin:2014aa,Balogh:2016aa}.
Specifically, at $z=1.05$, \citet{Muzzin:2014aa} finds $t_Q=1.00\pm0.25$ Gyr, the present work finds \gresult~Gyr, and \citet{Balogh:2016aa} finds $t_Q=1.5 \pm 0.5$ Gyr.
The results of \citet{Balogh:2016aa} are consistent within error bars with the present work, and \citet{Muzzin:2014aa} very nearly so.
Differences can be attributed to different approaches to measuring $t_Q$, including the above mentioned field corrections.
The definition of quenched galaxies differs as well, where \citet{Muzzin:2014aa} studies quenched poststarburst galaxies identified by their spectral features, \citet{Balogh:2016aa} uses an optical-IR color-color cut, and the present work uses a dust-corrected color-magnitude criterion.
Nevertheless, these three data points clearly indicate a quenching time between 1 and 1.5~Gyr.

\subsection{Redshift Evolution of Characteristic Timescales}

A clear evolutionary trend emerges from the assembled data points of Figure \ref{fig-tq}.
The quenching timescale at low redshift is long, roughly $4\mhyphen5$~Gyr, but has decreased to the order of $\sim1\mhyphen2$~Gyr at $z\sim1.5$.

Galaxy quenching may be the result of factors internal or external to the galaxy.
The former case includes scenarios where quenching occurs as a galaxy exhausts its gas reservoir (as in starvation, or overconsumption).
The latter case describes scenarios where quenching is due to the interaction of a galaxy with the host halo's environment at the high speeds typical of orbits within clusters.
In this section, we will endeavor to model several timescales associated with either gas depletion or kinematic effects, and plot them on Figure \ref{fig-tq}.

In gas depletion scenarios, the environment simply prevents cosmological accretion of fresh gas onto the galaxy, and what gas reservoir remains after infall is consumed by the galaxy over a gas depletion timescale $t_{\mathrm{depl}} = M_{gas}/\dot{M_{gas}}$, after which star formation ceases.
\citet{Fillingham:2015aa} note that measured molecular gas depletion timescales $t_{\mathrm{depl}}(\mathrm{H_{mol}})$ are much shorter than measured values of $t_Q$, over a broad range of redshifts.
This trend continues to be seen with the quenching timescales measured since the time of that study, including those in the present work.
In the local universe, however, \citet{Fillingham:2015aa} find very good agreement between the total gas depletion timescale $t_{\mathrm{depl}}(\mathrm{H_I}+\mathrm{H_{mol}})$ and the quenching times of high-mass galaxies ($M_* \geq 10^{9}~ \mathrm{M}_\odot$).
The first hypothesis we will consider is that the quenching timescale is simply the total gas depletion timescale,
where the galaxy's star-forming gas reservoir includes both atomic and molecular gas components.

As there are few observational constraints on galaxy atomic gas budgets at high redshift, we can only roughly approximate the redshift evolution of $t_{\mathrm{depl}}(\mathrm{H_I}+\mathrm{H_{mol}})$.
A star-forming galaxy's molecular gas fraction is found to decrease slowly with redshift out to $z=2$, by roughly a factor of 2 \citep{Genzel:2015aa,Tacconi:2017aa}, while the atomic gas density remains almost constant \citep{Bauermeister:2010aa}.
Since in the local universe, $M_{HI}\sim3M_{mol}$ \citep[see, e.g.,][]{Saintonge:2011aa}, for simplicity we will take $t_{\mathrm{depl}}(\mathrm{H_I}+\mathrm{H_{mol}})\sim~4~ t_{\mathrm{depl}}(\mathrm{H_{mol}})$, assuming the redshift evolution of $t_{\mathrm{depl}}(\mathrm{H_{mol}})$ from \citet[][equation 5]{Tacconi:2017aa}, and plot it on Figure \ref{fig-tq} (solid red line).

If galaxies experience significant star-formation-driven outflows, then the gas depletion timescale will be much shorter.
\citet{McGee:2014aa} has constructed a model parametrized by the ``mass-loading factor" $\eta$, such that the rate of gas mass ejected by a galaxy is a factor $\eta$ of the star formation rate.
We include on Figure \ref{fig-tq} the gas depletion time with outflows of $\eta=2.5$, using the cosmic evolution of the star formation rate derived by \citet{Whitaker:2012aa}.
This value of $\eta$ was found to best fit the quenching timescales described by \citet{McGee:2014aa}, and produces timescales that match $t_Q$ in clusters at low redshift.
While \citet{McGee:2014aa} intend for this timescale to model the delay time rather than the full quenching time, we include it on Figure \ref{fig-tq} to indicate its evolution with redshift (dotted orange line).
It is broadly the case that outflow timescales for various values of $\eta$ scale with redshift approximately as SFR, and so we also include on Figure \ref{fig-tq} the SFR evolution of \citet{Whitaker:2012aa}, normalized to a low-redshift timescale of 5 Gyr (dotted blue line).

If quenching is driven by gas stripping, $t_Q$ is expected to evolve as the dynamical time $t_{\mathrm{dyn}}$.
This dynamical time is commonly used to characterize timescales that depend on the kinematics of a galaxy within a cluster.
A cluster halo in virial equilibrium is characterized by relations between its radius $R$ and the velocity $V$ of its constituent galaxies, defining a dynamical timescale, $t_{\mathrm{dyn}}=R/V$.
From considerations of cosmology, the dynamical time is expected to scale with redshift as $t_{\mathrm{dyn}} \propto (1+z)^{-1.5}$.
If quenching is accomplished after a galaxy makes one or multiple passes through a particular radius of its host halo, $t_Q$ will be proportional to $t_{\mathrm{dyn}}$.
We normalize the dynamical timescale at low redshift separately to the SDSS group and cluster $t_Q$ data points.
We choose a normalization of $5.0\pm0.5$ Gyr for the cluster dynamical time scale, to span the two values for this data set reported by \citet{Wetzel:2013aa} and \citet{Balogh:2016aa}.
We normalize the group dynamical time scale to the 7 Gyr $t_Q$ reported by \citet{Balogh:2016aa}.
We plot these dynamical timescales also on Figure \ref{fig-tq} (solid and dashed green lines, respectively).

These trend lines roughly depict the expected evolution of $t_Q$ for various possible quenching scenarios.
They assume that the dominant quenching mechanism remains unchanged from low redshift, and is invariant for a given star formation rate and stellar mass.
We don't intend for these timescales to conclusively identify the mechanism responsible for environmental quenching, but rather to test if the measured redshift evolution of $t_Q$ is consistent with these possible models.

\subsection{Interpreting the Quenching Timescale}

The quenching timescale of massive galaxies ($M_* \geq 10^{10.5}~ \mathrm{M}_\odot$) is systematically higher in groups than in clusters.
In the SDSS sample at $z\sim0$, this trend is particularly pronounced, with $t_Q$ being higher in groups by $\sim2$ Gyr \citep{Balogh:2016aa}, although a difference is seen at all measured redshifts.
This difference cannot be entirely attributed to differences in background subtraction (see Section \ref{sec-systematics}), as demonstrated by the agreement between the present work and \citet{Balogh:2016aa} for the GCLASS cluster sample.
If $t_Q$ truly exhibits a dependence upon the mass of the host halo, then the quenching timescale is driven in part by factors external to the galaxy.

Referring to Figure \ref{fig-tq}, it is apparent that both estimates of an SFR-outflow timescale evolve too quickly at high redshift, and models with fixed mass-loading factor $\eta$ cannot simultaneously fit both the high- and low-redshift data points.

\citet{Balogh:2016aa} find that SFR-outflow quenching is a good fit to the delay times measured in the GCLASS and GEEC2 samples at $z\sim1$.
This conclusion is based in part on the quenching timescales measured in galaxies with masses $M_* \leq 10^{10.3}~ \mathrm{M}_\odot$, which we do not study here.
For those galaxies, $t_Q$ is found to be longer by several Gyr, and to increase with decreasing galaxy mass, in a way that is well-modeled by SFR outflows with $1.0 \leq \eta \leq 2.0$, although the same model is a poor fit at low redshift.
\citet{Balogh:2016aa} report that the dynamical timescale evolution is a good fit to $t_Q$ in galaxies with $M_* \geq 10^{10.5}~ \mathrm{M}_\odot$, as also noted by others \citep{Tinker:2010aa,Mok:2014aa}.
No disagreement is found between the present work and \citet{Balogh:2016aa}, where these studies overlap.

The cluster data points and group data points both evolve in accordance with the appropriately-normalized dynamical timescale.
The evolution of the dynamical time represents an evolution in the properties of groups and clusters (velocity dispersions, halo masses, etc.), not galaxy properties (SFR, gas fractions, etc.).
If quenching tracks $t_\mathrm{dyn}$, then it must be determined by the dynamical properties of clusters.
Such a scenario is often interpreted as being evidential of dynamical quenching scenarios such as ram-pressure stripping.

The estimated total gas depletion timescale $t_{\mathrm{depl}}(\mathrm{H_I}~+~\mathrm{H_{mol}})$ is a good fit for the quenching time at low redshift.
In the local universe, \citet{Fillingham:2015aa} also find that $t_Q$ for galaxies with masses $M_*~\geq~10^{10.5}~\mathrm{M}_\odot$ is well-fit by the total gas depletion timescale, $t_{\mathrm{depl}}(\mathrm{H_I}~+~\mathrm{H_{mol}})$.
Although we estimate $t_{\mathrm{depl}}(\mathrm{H_I}~+~\mathrm{H_{mol}})$ only very approximately from $t_{\mathrm{depl}}(\mathrm{H_{mol}})$ \citep[][equation 5]{Tacconi:2017aa}, we arrive at the same conclusion for low redshift clusters.

At high redshift, the estimated $t_{\mathrm{depl}}(\mathrm{H_I}~+~\mathrm{H_{mol}})$ does not evolve quickly enough to match $t_Q$.
Our estimate of the total gas depletion time assumes a molecular-to-atomic gas ratio that is unchanged from low redshift.
The atomic gas component of galaxies is poorly-constrained at high redshifts, and so the total depletion time we compare with here is a very rough extrapolation from the properties of low-redshift galaxies \citep[see discussion in ][]{Bauermeister:2010aa}.
Future work may better characterize the evolution of the total gas depletion timescale, and present estimates are not sufficient to rule out gas-consumption scenarios.

\section{Conclusions}\label{sec-conclusion}

In this paper we measured numbers of star-forming, intermediate, and quenched cluster members in two samples of galaxy clusters at $0.85 < z < 1.35$ and $1.35 < z < 1.65$.
A model of environmental quenching allows these number counts to constrain the quenching timescale $t_Q$.
From the analysis presented in this work, we draw the following conclusions:

\begin{itemize}

\item
We measure a quenching timescale of $t_Q= $\gresult{} Gyr in a sample of 10 galaxy clusters at $0.85 < z < 1.35$, and $t_Q=$\hiresult Gyr in a sample of 4 galaxy clusters at $1.35 < z < 1.65$.

\item
The evolution of the quenching timescale in clusters from the local universe to $z=1.55$ evolves faster than the molecular gas depletion timescale but slower than an SFR outflow model.
Instead, it appears to scale with the dynamical time, when normalized to the quenching timescale in local galaxy clusters.
This suggests that kinematical quenching mechanisms such as ram-pressure stripping may dominate in galaxies with masses $M_* \geq 10^{10.5}~ \mathrm{M}_\odot$ in clusters at high redshift, although we cannot rule out gas-depletion scenarios.

\item
The quenching timescale for galaxies with masses $M_* \geq 10^{10.5}~ \mathrm{M}_\odot$, measured out to $z\sim1.55$, appears to be shorter in clusters than in groups.
This indicates that environmental quenching mechanisms for these galaxies may depend on host halo mass at high redshift, as would be the case for kinematical quenching mechanisms such as ram-pressure stripping.

\end{itemize}

\acknowledgments{We thank Michael Balogh for his advice and contributions to this work prior to its publication.
We thank Matteo Fossati for providing the measurements shown in Figure 5.
G.W. acknowledges financial support for this work from NSF grant AST-1517863 and from NASA through programs GO-13306, GO- 13677, GO-13747 \& GO-13845/14327 from the Space Telescope Science Institute, which is operated by AURA, Inc., under NASA contract NAS 5-26555, and grant number 80NSSC17K0019 issued through the Astrophysics Data Analysis Program (ADAP).
JN is supported by Universidad Andres Bello internal research grant DI-18-17/RG.
Support for MCC was provided in part by NSF grant AST-1518257 as well as by NASA through grants AR-13242 and AR-14289 from the Space Telescope Science Institute, which is operated by the Association of Universities for Research in Astronomy, Inc., under NASA contract NAS 5-26555.
PC acknowledges the support of the FONDECYT postdoctoral research grant no 3160375.
Data presented herein were obtained using the UCI Remote Observing Facility, made possible by a generous gift from John and Ruth Ann Evans.
Some of the data presented herein were obtained at the W. M. Keck Observatory, which is operated as a scientific partnership among the California Institute of Technology, the University of California and the National Aeronautics and Space Administration.
The Observatory was made possible by the generous financial support of the W. M. Keck Foundation.
The authors wish to recognize and acknowledge the very significant cultural role and reverence that the summit of Maunakea has always had within the indigenous Hawaiian community.
We are most fortunate to have the opportunity to conduct observations from this mountain.}

\appendix

\section{Environmental Quenching Model}\label{sec-math}

\begin{figure}[h!]
\centering \includegraphics[width=0.5\textwidth]{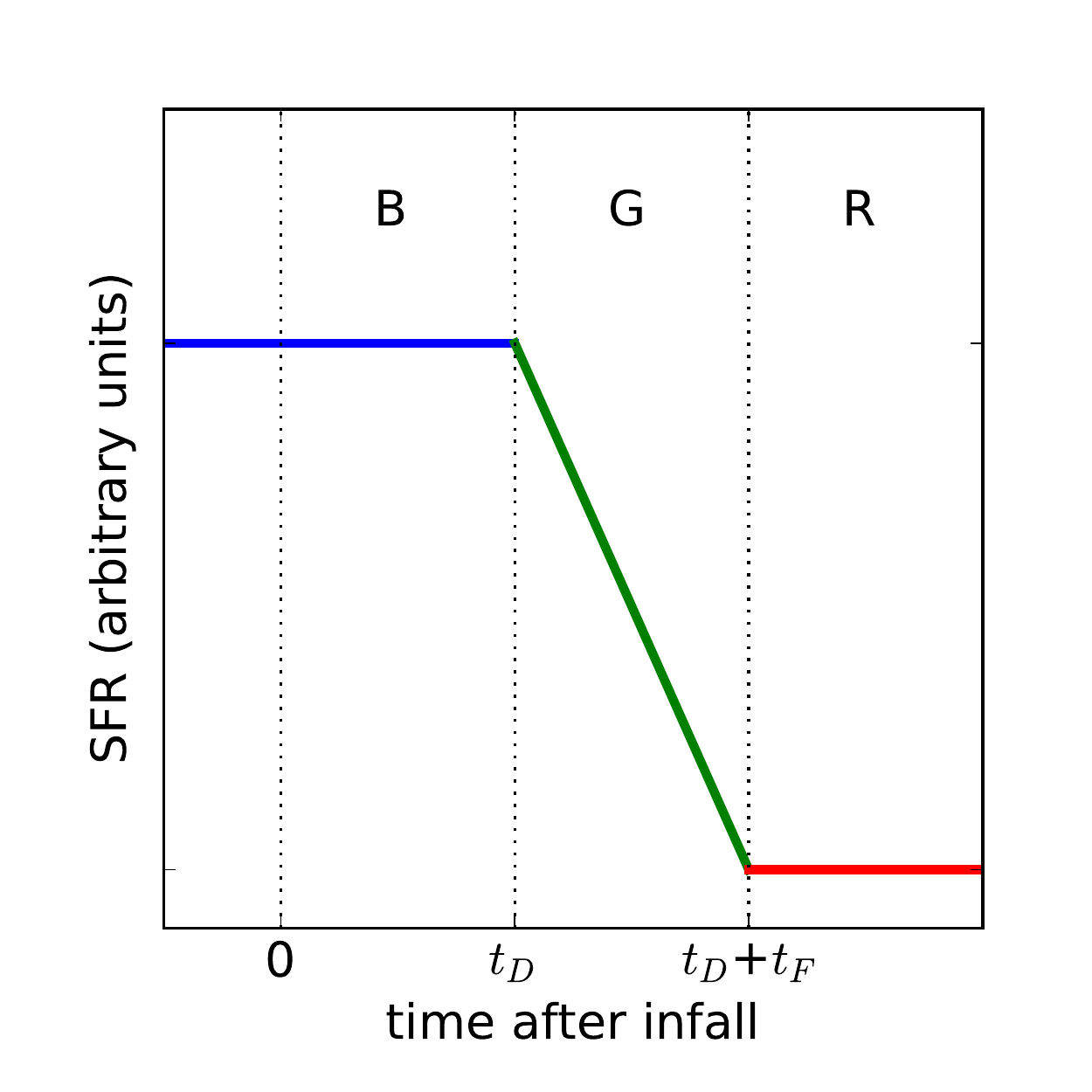}
\caption{Model star formation rate of a galaxy as a function of time relative to its accretion by the cluster.
The galaxy's color reflects its star formation rate, such that star-forming galaxies are labeled blue, galaxies with declining star formation rate are labeled green, and quiescent galaxies are labeled red.
All galaxies that fall into the cluster are assumed to be star-forming, and remain star-forming for a delay time $t_D$.
Following the delay period, star formation begins to quench over a fade time, $t_F$, after which the galaxy is quiescent.
The total quenching time $t_Q$ is $t_D + t_F$.
\label{fig-a-sfh}}
\end{figure}

In this model, environmental quenching is characterized by a quenching timescale $t_Q$, defined as the length of time after accretion for a galaxy to be completely quenched.
A galaxy's time in the cluster is divided into three evolutionary phases: a (blue) delay phase, wherein star formation continues as if unaffected by environment, a (green) fade phase, during which star formation declines, and a (red) quenched phase, after star formation has fully ceased.
The observed colors of galaxies trace their star formation rate and therefore the galaxy's evolutionary phase (see Figure \ref{fig-ellipses}), and form the basis for labeling the delay, fade, and quenched phases as blue, green, and red, respectively.

We take as given the observed numbers of red, green, and blue galaxies in a cluster (R, G, and B, respectively), at the cluster redshift, $z_c$.
For our purposes, it is first necessary to correct for galaxies that were already quenched before they fell into the cluster.
We first calculate the total number of quiescent galaxies accreted from the field over the lifetime of the cluster using the field galaxy mass functions computed by \citet{Muzzin:2013ab}.
We then subtract this number from the observed number of red galaxies, leaving only galaxies that were blue when accreted by the cluster.
This field-quenched correction is described in detail in \ref{sec-a-field}.
For the rest of this discussion, we assume corrected number counts of galaxies, and that these galaxies were star-forming when accreted.

A (blue) star-forming galaxy that falls into the cluster will remain star-forming for a delay time, $t_D$.
After the passage of one delay time $t_D$, the galaxy's star formation rate fades over the fade time, $t_F$.
Subsequent to a total amount of time $t_Q = t_D + t_F$, a galaxy has completely ceased forming stars, and is considered quiescent.
In Figure \ref{fig-a-sfh}, we show this evolution of galaxy type schematically as a function of time following infall.

From this, it follows that star-forming (blue) cluster members were accreted as recently as up to one $t_D$ ago, and so are still in their star-forming ``delay" phase.
Intermediate (green) cluster members, in the ``fade" phase, were accreted between $t_D$ and $t_D + t_F$ ago.
Quenched (red) cluster members are all galaxies accreted earlier than that.
The quenching time $t_Q$ is then the sum of the delay time, $t_D$, and a fade time, $t_F$.

The central assumption of this model is that all galaxies undergo the same evolutionary process, passing from blue to green to red once accreted by the cluster.
Because of this, the numbers of blue and green galaxies found in the cluster trace the amount of time spent in the delay and fade phases of evolution, and red galaxies trace the integrated history of all galaxy accretion older than one quenching time.

Given a galaxy accretion rate $\mathrm{d}N/\mathrm{d}t$, the numbers of red, green, and blue galaxies can constrain the times $t_D$ and $t_F$.
Specifically,

\begin{empheq}{align*}
&B = \int_{-t_D}^{0} \mathrm{d}N/\mathrm{d}t\ \mathrm{d}t\\
&G = \int_{-(t_D+t_F)}^{-t_D} \mathrm{d}N/\mathrm{d}t\ \mathrm{d}t\\
&R = \int_{-t_H}^{-(t_D+t_F)} \mathrm{d}N /\mathrm{d}t\ \mathrm{d}t
\end{empheq}

where B, G, and R are the numbers of blue, green, and red cluster galaxies, respectively, observed at time $t=0$, and $t_H$ is the Hubble time.
Note that the negative sign of the integration limits emphasizes the fact that these galaxies were accreted in the past.
While we have begun by stating functions here in terms of time $t$ relative to the cluster, later we will cast our equations in terms of redshift for easier use with real data.

In principle, the galaxy accretion rate $\mathrm{d}N/\mathrm{d}t$ is some fraction of the total halo mass accretion rate $\mathrm{d}M/\mathrm{d}t$, determined by the baryon and gas fractions of galaxies, and related to observed counts by the stellar mass function above the mass completeness of our sample.
However, it is not necessary to calculate this factor if we consider ratios of galaxy counts instead of absolute numbers.
Given that galaxy stellar mass is some fraction of the mass accreted by the cluster, such that $\mathrm{d}N/\mathrm{d}t\ = f_G\ \mathrm{d}M/\mathrm{d}t$, it follows that

\begin{equation*}
\frac{\displaystyle\int_{t_2}^{t_1} \mathrm{d}N/dt\ \mathrm{d}t}{\displaystyle\int_{t_3}^{t_2} \mathrm{d}N/\mathrm{d}t\ dt} = \frac{\displaystyle\int_{t_2}^{t_1} f_G\ \mathrm{d}M/\mathrm{d}t\ \mathrm{d}t}{\displaystyle\int_{t_3}^{t_2} f_G\ \mathrm{d}M/\mathrm{d}t\ \mathrm{d}t}
\end{equation*}

for arbitrary times $t_1$, $t_2$, $t_3$.
If we assume $f_G$ remains relatively constant with time, we can cancel it from the right-hand side of the above equation, and can therefore express ratios of galaxy counts purely in terms of the mass accretion rate, $\mathrm{d}M/\mathrm{d}t$.

Cosmological N-body simulations can make predictions for the mass accretion histories of cluster-scale dark matter halos \citep{Lacey:1993aa}.
\citet{Fakhouri:2010aa} has used merger histories in the Millennium-\textsc{II} simulation to fit an expression for the mean mass growth rate of halos of the form

\begin{equation*}
\frac{\mathrm{d}M}{\mathrm{d}t} = 46.1\ \mathrm{M_\odot\ yr^{-1}} \left(\frac{M}{10^{12}\ \mathrm{M_\odot}}\right)^{1.1}\times(1+1.11z)\sqrt{\Omega_{\mathrm{m}}(1+z)^3 + \Omega_{\Lambda}}
\end{equation*}

for a halo of mass $M$ at redshift $z$.

A change of units yields

\begin{empheq}{align*}
&\frac{\mathrm{d}M}{\mathrm{d}z} = \frac{-t_H}{46.1\ \mathrm{yr}} \times \left(\frac{1+1.11z}{1+z}\right)\left(\frac{M}{10^{12}\ \mathrm{M_\odot}}\right)^{1.1} \mathrm{M_\odot}\\
&M(z=1.6) = 3\times10^{14}\ \mathrm{M_\odot}\\
\end{empheq}

where we have used the estimated mean cluster mass of the $z=1.6$ cluster sample as a boundary condition (Wilson 2018, in prep).
When calculating quenching timescales for the lower-redshift cluster sample, the mean cluster mass boundary condition is $M=3.8\times10^{14}\ \mathrm{M_\odot}$ at $z=1$.
We note that our $z=1.6$ cluster sample has a mean halo mass that is only slightly higher than that of progenitors of the $z=1$ sample \citep{Lidman:2012aa,Nantais:2017aa}, and our results do not depend strongly on the choice of host halo mass for a reasonable range of masses.

\begin{figure}
\centering \includegraphics{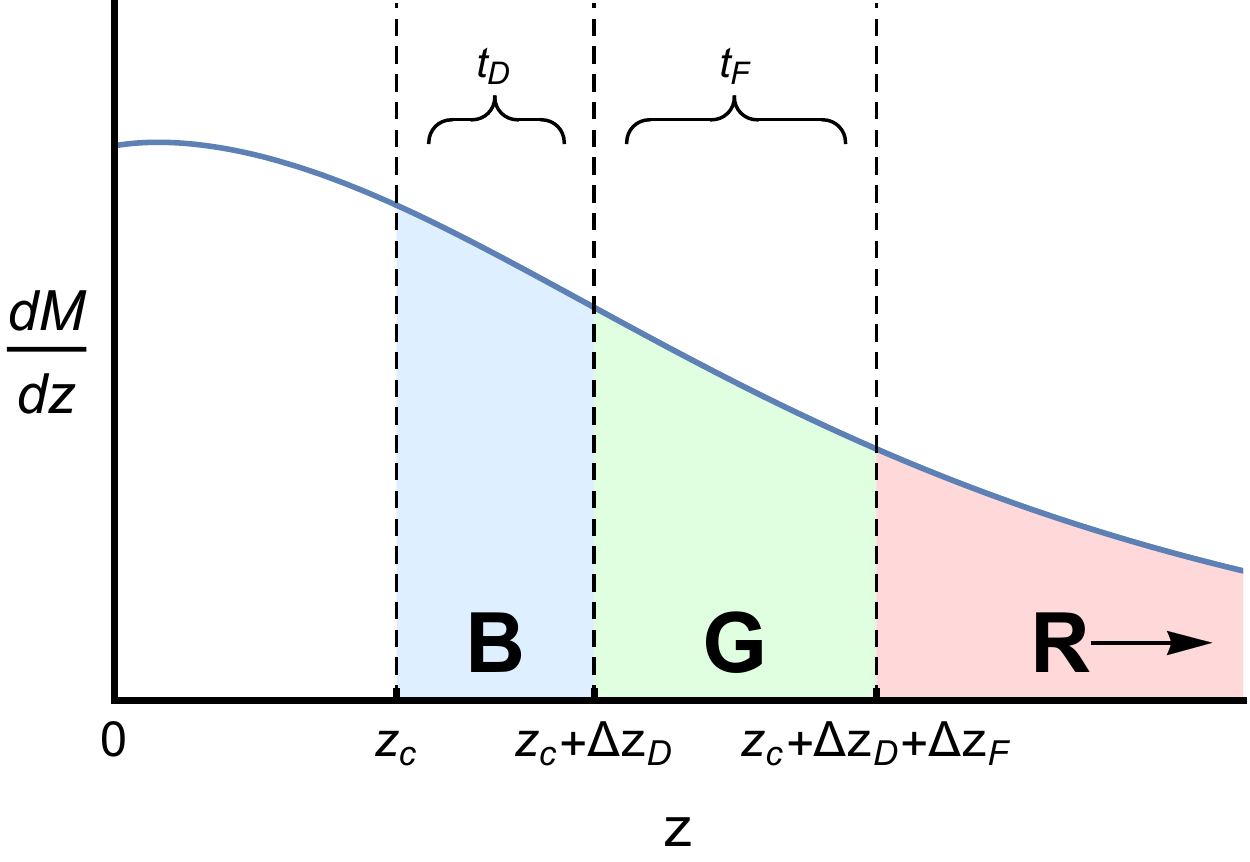}
\caption{Cluster mass accretion rate $\mathrm{d}M/\mathrm{d}z$ as a function of redshift, for a cluster observed at redshift $z_c$.
The number of galaxies accreted over a given redshift interval is proportional to the area under the curve for that interval.
Blue galaxies, being accreted no later than one $t_D$ ago, have numbers proportional to the integral of the mass accretion rate between $z_c$ and $z_c + \Delta z_D$, labeled B.
Green galaxies have been in the cluster longer than one $t_D$ but no longer than $t_D + t_F$ and so have been accreted over the interval between $z_c + \Delta z_D$ and $z_c + \Delta z_D + \Delta z_F$, labeled G.
The number of red galaxies, R, is proportional to the integral of all mass accretion that occurred at redshifts greater than $z_c + \Delta z_D + \Delta z_F$.
\label{fig-rgb}}
\end{figure}

This system of equations can be solved numerically for $M(z)$, the total cluster mass as a function of redshift, and $\mathrm{d}M/\mathrm{d}z$, the mass accretion rate.
By recasting our earlier set of equations to be functions of redshift, we can now write

\begin{align*}
&\frac{B}{G+R} = \frac{\displaystyle\int_{z_c + \Delta z_D}^{z_c} \mathrm{d}M/\mathrm{d}z\ \mathrm{d}z}{\displaystyle\int_{\infty}^{z_c + \Delta z_D} \mathrm{d}M/\mathrm{d}z\ \mathrm{d}z}\\
&\frac{G}{R} = \frac{\displaystyle\int_{z_c + \Delta z_D + \Delta z_F}^{z_c + \Delta z_D} \mathrm{d}M/\mathrm{d}z\ \mathrm{d}z}{\displaystyle\int_{\infty}^{z_c + \Delta z_D + \Delta z_F} \mathrm{d}M/\mathrm{d}z\ \mathrm{d}z}
\end{align*}

where, for a cluster at $z=z_c$, $z_c + \Delta z_D$ is the redshift one delay time $t_D$ ago, and $z_c + \Delta z_D + \Delta z_F$ is one delay plus fade time, $t_D + t_F$, ago.
The relationship between these variables is summarized visually in Figure \ref{fig-rgb}.

With an expression for $M(z)$, the integral relations become

\begin{align*}
&\frac{B}{G+R} = \frac{M(z_c) - M(z_c + \Delta z_D)}{M(z_c + \Delta z_D)}\\
&\frac{G}{R} = \frac{M(z_c + \Delta z_D) - M(z_c + \Delta z_D + \Delta z_F)}{M(z_c + \Delta z_D + \Delta z_F)}
\end{align*}

where we have used the fact that $M(z)=0$ when $z\to\infty$.

Altogether we apply the following set of three equations with three unknowns, and one boundary condition:

\begin{empheq}{align}
&\frac{\mathrm{d}M}{\mathrm{d}z} = \frac{-t_H}{46.1\ \mathrm{yr}} \times \left(\frac{1+1.11z}{1+z}\right)\left(\frac{M}{10^{12}\ \mathrm{M_\odot}}\right)^{1.1} \mathrm{M_\odot} \label{eq-model-first} \\
&\frac{B}{G+R} = \frac{M(z_c) - M(z_c + \Delta z_D)}{M(z_c + \Delta z_D)} \label{eq-model-zd} \\
&\frac{B+G}{R} = \frac{M(z_c + \Delta z_D) - M(z_c + \Delta z_D + \Delta z_F)}{M(z_c + \Delta z_D + \Delta z_F)} \label{eq-model-zf}\\
&M(1.6) = 3\times10^{14}\ \mathrm{M_\odot} \label{eq-model-last}
\end{empheq}

\begin{figure}
\centering \includegraphics[width=\textwidth]{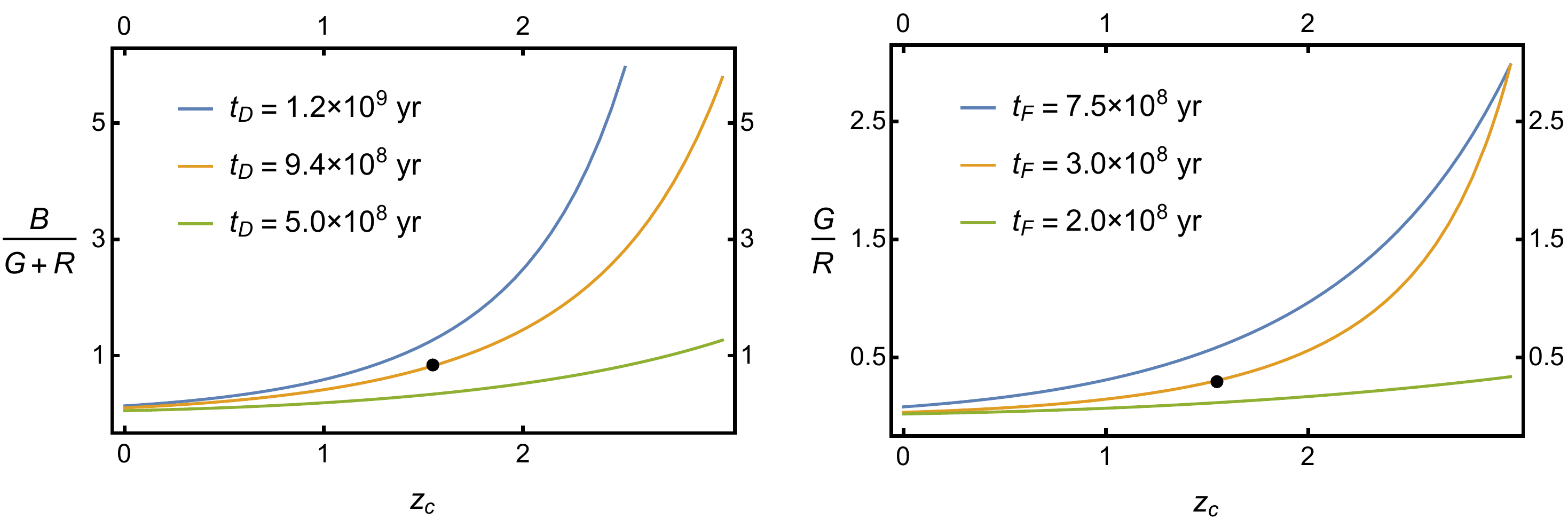}
\caption{Modeled evolution of the fractions $\frac{B}{G+R}$ and $\frac{G}{R}$.
Lines show the evolution of these fractions for the indicated delay and fade times, $t_D$ and $t_F$.
Note that the fraction of blue galaxies increases with redshift, and with longer delay times, as expected.
The black point indicates the measured value of these fractions for the stacked high-redshift sample, at the mean redshift of the sample, $z_c = 1.55$.
From the left panel, it is clear that a delay time of $t_D = 9.4\times10^8$ yr is indicated in order to produce the observed fraction of blue galaxies.
With this value for $t_D$ we plot the redshift evolution of $\frac{G}{R}$ in the right panel, given that green galaxies were accreted between $t_D$ and $t_F$ ago, for selected values of $t_F$.
A value of $3.0\times10^8$ yr is indicated for $t_F$.
\label{fig-frac}}
\end{figure}

Through Equations~\eqref{eq-model-first}~--~\eqref{eq-model-last}, the numbers of red, green, and blue galaxies at cluster redshift $z_c$ constrain the delay and fade redshift intervals, $\Delta z_D$ and $\Delta z_F$.
For our purposes, we find it easiest to first solve the differential equation for $M(z)$ numerically with Mathematica using \texttt{NDSolve}.
Knowing $M(z)$, it is then a matter of finding the redshift interval $\Delta z_D$ that satisfies Equation \ref{eq-model-zd}, which we accomplish with \texttt{FindRoot}.
We repeat the process to then determine $\Delta z_F$ from Equation \ref{eq-model-zf}.

To illustrate the method, we plot the modeled evolution of the fractions $\frac{B}{G+R}$ and $\frac{G}{R}$ in Figure \ref{fig-frac} for selected values of $t_D$ and $t_F$.
From this plot, it is clear that the observed ratios of red, green, and blue galaxies constrain $t_D$ and $t_F$.

Having determined $\Delta z_D$ and $\Delta z_F$, we can apply standard cosmology to calculate the time intervals

\begin{align*}
&t_D = t_H \int_{z_c}^{z_c+\Delta z_D}\frac{\mathrm{d}z}{(1+z)\sqrt{\Omega_m(1+z)^3 + \Omega_\Lambda}}\\
&t_F = t_H \int_{z_c+\Delta z_D}^{z_c+\Delta z_D+\Delta z_F}\frac{\mathrm{d}z}{(1+z)\sqrt{\Omega_m(1+z)^3 + \Omega_\Lambda}}
\end{align*}

and thereby measure the quenching timescale, $t_Q = t_D + t_F$.

The technique we describe here relies on interpreting the integrated mass accretion history of a cluster, and so the resulting quenching timescales are time-averaged over the history of the cluster.
This should not impact the results of this paper as the clusters we study here are still very young, but would need to be taken into consideration when applying this technique at low redshift.

\subsection{Field-quenched Correction}\label{sec-a-field}

Quenched galaxies exist in the field, and therefore some of the galaxies accreted by a cluster will already be quenched.
If these galaxies are included when calculating $t_Q$, they will inflate the relative proportion of red galaxies, resulting in an apparently shorter quenching time.
Correcting for this is a simple matter of calculating the fraction of galaxies that were quiescent when accreted, and subtracting them from the total number of red galaxies.

We start by calculating the quiescent fraction of field galaxies above the mass completeness limit of $10^{10.5}~ \mathrm{M}_\odot$ as a function of redshift, $f_Q(z)$.
\citet{Muzzin:2013ab} provides Schechter mass function fits to field galaxies in the COSMOS/UltraVISTA survey.
These functions have the form

\begin{equation*}
\Phi(M) = \mathrm{ln}\ 10\times\Phi^*\times10^{(M-M^*)(1+\alpha)}\times \mathrm{exp}(-10^{(M-M^*)})
\end{equation*}

and are parametrized by a normalization, $\Phi^*$, a characteristic mass, $M^*$, and a low-mass-end slope, $\alpha$.
The masses $M$ and $M^*$ are logarithmic stellar masses of the form $M=\mathrm{log}_{10}(M_{star}/M_\odot)$.
\citet[][Table 1]{Muzzin:2013ab} fits separate mass functions for star-forming and quiescent galaxies in seven redshift bins from $0.2 \leq z \leq 4.0$

From these mass functions we can define the field quiescent fraction $f_Q(z_i)$ at seven redshifts $z_i$,

\begin{equation*}
f_Q(z_i) = \frac{\int_{10.5}^{\infty}~\Phi_Q(M, z_i)~\mathrm{d}M}{\int_{10.5}^{\infty}~\Phi_Q(M, z_i)~\mathrm{d}M+\int_{10.5}^{\infty}~\Phi_A(M, z_i)~\mathrm{d}M}
\end{equation*}

where $\Phi_Q(M, z_i)$ and $\Phi_A(M, z_i)$ are the quiescent and star-forming mass functions, respectively, and $z_i$ is the mean redshift of the $i^{\mathrm{th}}$ redshift bin.

The fraction of quiescent field galaxies with masses above $10^{10.5}~ \mathrm{M}_\odot$ evolves with cosmic time as the cluster accretes galaxies from the field.
To determine the total fraction of quiescent field galaxies accreted over the lifetime of the cluster, we must integrate the galaxy accretion rate weighted by the field quiescent fraction.
Therefore we interpolate $f_Q(z_i)$ between the seven redshift points by fitting $3^{\mathrm{rd}}$-order polynomial curves between successive data points using the Mathematica function \texttt{Interpolation}.
By default this creates a continuous and differentiable 3\textsuperscript{rd}-order polynomial function $f_Q(z)$ suitable for integration.

\begin{figure}
\centering \includegraphics[width=0.5\textwidth]{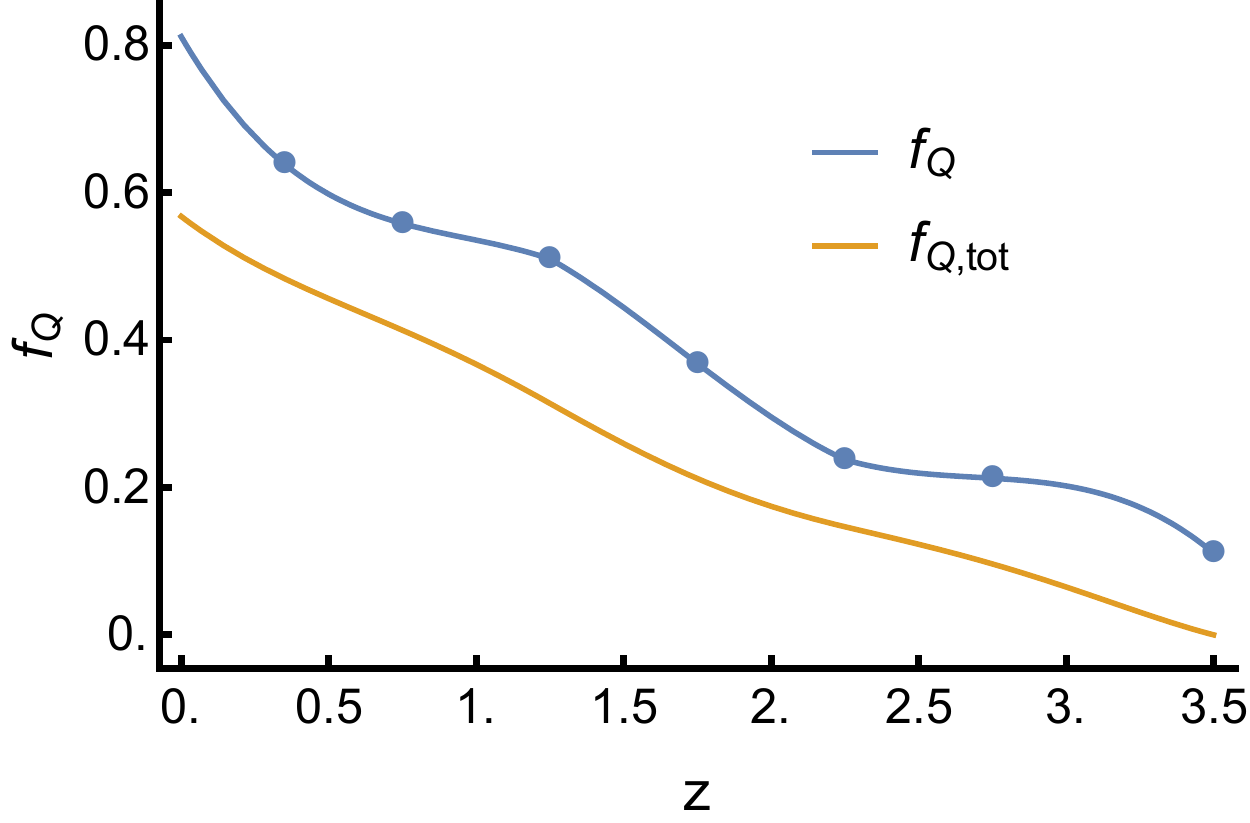}
\caption{Evolution of the field quiescent fraction with redshift.
The field quiescent fraction is determined from the field mass function fits of \citet{Muzzin:2013ab} in seven redshift bins, for galaxies with masses $M \geq 10^{10.5}~ \mathrm{M}_\odot$, plotted as points.
The blue line depicts a function interpolated from the seven points.
The orange line is the integrated mass accretion rate of a cluster weighted by the field quiescent fraction, or the total fraction of accreted quiescent field galaxies.
\label{fig-a-schechter}}
\end{figure}

Previously, we used the cluster mass accretion rate, $\mathrm{d}M/\mathrm{d}z$, as a proxy for the cluster galaxy accretion rate.
The total accreted field quiescent fraction $f_\mathrm{Q,tot}(z)$ is therefore

\begin{equation}\label{eq-ftot}
f_\mathrm{Q,tot}(z) = \frac{\int_{-\infty}^z~ \mathrm{d}M/\mathrm{d}z'~f_Q(z')~\mathrm{d}z'}{\int_{-\infty}^z~ \mathrm{d}M/\mathrm{d}z'~\mathrm{d}z'}.
\end{equation}

where $z$ is the redshift of the cluster.
The evolution of $f_\mathrm{Q,tot}(z)$ and $f_Q(z)$ with redshift is shown in Figure \ref{fig-a-schechter}.

From Equation \ref{eq-ftot}, we can determine the fraction of quiescent galaxies in a cluster at redshift $z$ that were already quenched at the time they were accreted.
We therefore multiply the number of red galaxies in each cluster by $1-f_\mathrm{Q,tot}(z_c)$ before applying Equations~\eqref{eq-model-first}~--~\eqref{eq-model-last} and determining $t_Q$.

\section{Classifying star-forming and quiescent galaxies with a UVJ method}\label{sec-a-uvj}

\begin{figure}
\centering \includegraphics[width=\textwidth]{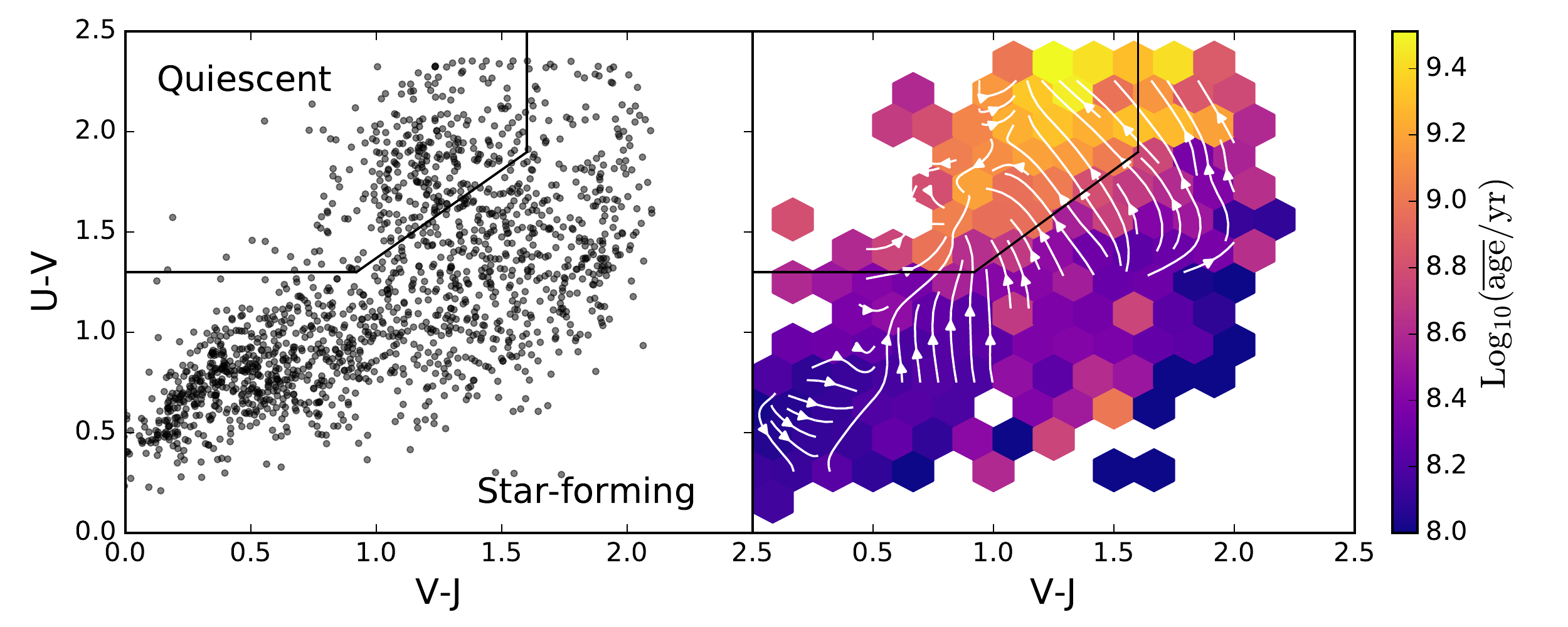}
\caption{Left panel: Rest-frame \textit{U-V} versus \textit{V-J} color-color diagram for all cluster members in the high-redshift sample.
Right panel: 2D histogram of mean binned galaxy ages in rest-frame \textit{UVJ} space.
The ages depicted here are derived from SED fitting (see Section \ref{sec-fast}).
The vector field plotted in white depicts the negative gradient of the mean binned ages, representing a possible approximation of evolutionary tracks.
Almost all of these tracks depict galaxies moving from the star-forming to the quiescent bin, and therefore represent quenching (intermediate) galaxies.
Note that these tracks take paths that cross all portions of the boundary between the star-forming and quiescent regions.
\label{fig-a-UVJ}}
\end{figure}

Rest-frame \textit{UVJ} color-color selection is frequently used to distinguish quiescent and star-forming galaxies, by dividing the space of rest-frame \textit{U-V} versus \textit{V-J} colors into a star-forming and a quiescent region.
The cuts that define these regions have been empirically derived by \citet{Williams:2009tt}, being tuned to maximally reflect the bimodality of galaxy populations out to $z\sim2.5$.
The \textit{UVJ} method accounts for dust reddening by using two colors that differ in their sensitivity to star formation and dust, to break the degeneracy between old-and-quiescent and star-forming-and-dusty galaxies.
In Figure \ref{fig-a-UVJ}, we plot the \textit{UVJ} color-color diagram for all cluster members in our sample.

The \textit{UVJ} method parallels the selection used in Section \ref{sec-ellipses} to classify quiescent (red), star-forming (blue), and intermediate (green) galaxies.
A natural question is whether similar values for $t_Q$ are obtained when galaxies are classified according to their \textit{UVJ} colors rather than the dust-corrected color-magnitude method.
In this appendix we will perform this comparison and report the results.
This subject will be expanded on in a letter (Foltz 2018, in prep).

Equations ~\eqref{eq-model-first}~--~\eqref{eq-model-last} are written in terms of the observed number of red, green, and blue galaxies in a cluster.
The \textit{UVJ} method (as it is commonly used) however only classifies galaxies as either star-forming or quiescent.
The principal difficulty in identifying an intermediate \textit{UVJ} region lies in the fact that a galaxy's location in \textit{UVJ} space is strongly dependent on both its star formation rate and its dust reddening.

For example, a galaxy in the upper-right region of the star-forming bin is both star-forming and very dust-reddened.
If it quenches, after some time it will end up in the quiescent bin, where star formation rates are low and dust-reddening is low.
The galaxy will need to decrease in dust-reddening as it decreases its star formation rate, and its precise trajectory in \textit{UVJ} space will depend on the details of how both of these values change in time.
The \textit{UVJ} green valley is therefore defined not only by intermediate star formation rates, but also by intermediate dust-reddening values.

This point is illustrated further in Figure \ref{fig-a-UVJ}.
The right panel of this figure depicts mean binned ages of galaxies in rest-frame \textit{UVJ} space, and the gradient of these mean ages is shown as a white vector field.
Intermediate galaxies, by definition, are those moving from the star-forming to the quiescent bin, and the age bins indicate many possible paths such galaxies might take as they age.
This makes it difficult to know where to look in \textit{UVJ} space for galaxies that have recently shut off their star formation, although it is natural to suppose that they must lie along the boundary of the quiescent and star-forming regions, especially since that boundary was drawn precisely to separate these two populations.
At the very least, there is reasonable doubt about the specific evolutionary tracks of quenching galaxies in a \textit{UVJ} diagram, due to the lack of a prescription for modeling how dust reddening will change following the cessation of star formation.
In contrast, the evolution of quenched galaxies in Figure \ref{fig-ellipses} is unambiguous, allowing a straightforward identification of blue, green, and red galaxies.

There have been some attempts to augment the \textit{UVJ} method with the addition of a third bin.
\citet{Whitaker:2012aa} subdivides the quiescent bin into young and old sections, in light of the fact that the color sequence of \textit{UVJ}-quiescent galaxies is driven by the ages of their stellar populations \citep{Whitaker:2010aa,Whitaker:2012aa}.
We wish to emphasize that this \textit{V-J} cut is successful for the purposes of \citet{Whitaker:2010aa,Whitaker:2012aa} in that it identifies young, quiescent galaxies.
We simply caution against others interpreting this cut as a general intermediate bin, as the age-color relation does not extend to the full population of galaxies, where the picture is complicated by dust reddening.
There is a difference between young quiescent galaxies and intermediate galaxies in general.
For the purposes of our quenching model, it is necessary to identify intermediate galaxies that have just left the star-forming blue cloud.

In a different approach, by adapting the method described in Appendix \ref{sec-math}, we can measure a quenching time using only numbers of star-forming and quiescent galaxies.
The general approach is to omit the number of intermediate galaxies (G) by assuming they are included in the number of star-forming galaxies (B), under the assumption that their declining but nonzero star formation rates will count them among the star-forming galaxies in the \textit{UVJ} diagram.
We then reformulate our equations under this assumption as follows:
the loss of the known variable G comes at the cost being unable to solve for $t_D$ and $t_F$ separately, and so we solve for $t_Q$ directly without separating it into delay and fade times.

Mathematically, if we apply the following transformation:

\begin{equation*}
\begin{aligned}
R'&=R    & t_D'&=t_D+t_F=t_Q\\
G'&=0    & t_F'&=0\\
B'&=B+G & \\
\end{aligned}
\end{equation*}

then the earlier integral relations simplify to

\begin{equation*}
\frac{B'}{R'} = \frac{\displaystyle\int_{z_c + \Delta z_Q}^{z_c} \mathrm{d}M/\mathrm{d}z\ \mathrm{d}z}{\displaystyle\int_{\infty}^{z_c + \Delta z_Q} \mathrm{d}M/\mathrm{d}z\ \mathrm{d}z}
\end{equation*}

where $\Delta z_Q$ is the redshift interval that spans one quenching time $t_Q$, $B'$ is the number of \textit{UVJ}-star-forming galaxies, and $R'$ is the number of \textit{UVJ}-quiescent galaxies.
From here, the arguments of Appendix \ref{sec-math} follow, and we can use the \textit{UVJ}-derived number counts to constrain a quenching time with the following set of equations:

\begin{align*}
&\frac{\mathrm{d}M}{\mathrm{d}z} = \frac{-t_H}{46.1\ \mathrm{yr}} \times \left(\frac{1+1.11z}{1+z}\right)\left(\frac{M}{10^{12}\ \mathrm{M_\odot}}\right)^{1.1} \mathrm{M_\odot}\\
&M(1.6) = 3\times10^{14}\ \mathrm{M_\odot}\\
&\frac{B}{R} = \frac{M(z_c) - M(z_c + \Delta z_Q)}{M(z_c + \Delta z_Q)}
\end{align*}

As in Section \ref{sec-results}, we stack each cluster sample by taking the total numbers of \textit{UVJ}-quiescent and \textit{UVJ}-star-forming galaxies at the mean redshift of both cluster samples.
These number counts then constrain a quenching timescale as described in Appendix \ref{sec-math}.
Poisson counting statistics and a Monte Carlo simulation with 200 iterations provides the 68\% confidence interval, as described in Section \ref{sec-error}.
The results are shown in Table \ref{tbl-UVJ}, alongside the results of the main analysis for comparison.
The quenching timescales derived by both methods very nearly agree within uncertainties.
For the GCLASS sample at $z=1.0$, we find $\tq{1.11}{0.16}{0.20}$~ Gyr, compared to $\simgresult$ Gyr for the RGB classification method.
In the higher-redshift sample at $z=1.55$, we find $\tq{1.16}{0.12}{0.14}$~ Gyr, compared to $\simhiresult$ Gyr for the RGB method.

Our error bars are likely under-estimated when adapting the Monte Carlo method to the UVJ classification, as it describes uncertainty in only two variables (RB) instead of the RGB method's full three.
The UVJ method yields a $t_Q$ that is lower in both cases because it finds a slightly higher passive fraction.
This is indicative of the way both classification schemes treat intermediate galaxies, which are necessarily split between the UVJ-star-forming and UVJ-quiescent categories.
Not all G galaxies are included in the UVJ-star-forming category, having instead been classified UVJ-quiescent.


\begin{deluxetable}{ccccccc}
\tabletypesize{\scriptsize}
\tablecaption{Effect of \textit{UVJ} selection on inferred quenching timescales\label{tbl-UVJ}}
\tablecolumns{9}
\tablewidth{0pt}
\tablehead{
\colhead{Cluster} \vspace{-0.4cm} & & & & & \colhead{UVJ ${t_Q}^a$} & \colhead{RGB ${t_Q}^b$}\\ \vspace{-0.4cm}
& \colhead{N} & \colhead{$\bar{z}$} & \colhead{Quiescent} & \colhead{Star-Forming} & & \\
\colhead{Sample} & \colhead{} & \colhead{} & \colhead{} & \colhead{} & \colhead{(Gyr)} & \colhead{(Gyr)}
}
\startdata
GCLASS & 10 & 1.04 & 187 & 58 & $\tq{1.11}{0.16}{0.20}$ & \gresult \\
SpARCS high-redshift & 4 & 1.55 & 85 & 75 & $\tq{1.16}{0.12}{0.14}$ & \hiresult \\
\enddata
\tablenotetext{a}{Quenching timescale derived using UVJ classification}
\tablenotetext{b}{Quenching timescale derived using dust-corrected U-B color-magnitude classification, for comparison   (see Section \ref{sec-ellipses})}
\end{deluxetable}


\clearpage
\bibliography{mybib}
\clearpage

\end{document}